\documentclass[11pt,a4paper]{article}
\usepackage{jheppub}
\usepackage{xcolor}
\usepackage{graphicx}
\usepackage{hyperref}
\usepackage[all]{hypcap}
\usepackage[utf8x]{inputenc}
\usepackage{listings}
\usepackage{xspace}
\usepackage{relsize}
\usepackage{multirow}
\usepackage{caption}
\usepackage{subcaption}

\hypersetup{
    unicode=false,          % non-Latin characters in Acrobat's bookmarks
    pdftoolbar=true,        % show Acrobat's toolbar?
    pdfmenubar=true,        % show Acrobat's menu?
    pdffitwindow=false,     % window fit to page when opened
    pdfstartview={FitH},    % fits the width of the page to the window
    pdftitle={HEPfit Manual},    % title
    pdfauthor={HEPfit Developers},     % author
    pdfsubject={HEPfit: a Code for the Combination of Indirect and Direct Constraints on High Energy Physics Models},   % subject of the document
    pdfcreator={HEPfit},    % creator of the document
    pdfproducer={HEPfit},   % producer of the document
    pdfkeywords={Flavour Physics} {Higgs Physics} {EW Physics} {Computational HEP}, % list of keywords
    pdfnewwindow=true,      % links in new window
    colorlinks=true,        % false: boxed links; true: colored links
    linktocpage=true,
}

\definecolor{mygreen}{rgb}{0,0.6,0}
\definecolor{mygray}{rgb}{0.5,0.5,0.5}
\definecolor{mymauve}{rgb}{0.58,0,0.82}
\DeclareMathAlphabet{\mathpzc}{OT1}{pzc}{m}{it}
\newcommand{\mathpzcB}[1]{\mathlarger{\mathlarger{\mathpzc{#1}}}}

\newcommand{\HEPfit}{\texttt{HEPfit}\xspace}
\newcommand{\BAT}{\texttt{BAT}\xspace}
\newcommand{\ROOT}{\texttt{ROOT}\xspace}

\title{\HEPfit: a Code for the Combination of Indirect and Direct
  Constraints on High Energy Physics Models}
  
\author[a,b]{J.~de Blas,}
\author[c,d]{D.~Chowdhury,}
\author[e]{M.~Ciuchini,}
\author[f]{A.~M.~Coutinho,}
\author[g]{O.~Eberhardt,}
\author[h]{M.~Fedele,}
\author[i]{E.~Franco,}
\author[j]{G.~Grilli di Cortona,}
\author[g]{V.~Miralles,}
\author[k]{S.~Mishima,}
\author[l,m]{A.~Paul,}
\author[g]{A.~Pe{\~n}uelas,}
\author[n]{M.~Pierini,}
\author[o]{L.~Reina,}
\author[i,p]{L.~Silvestrini,}
\author[q]{M.~Valli,}
\author[e]{R.~Watanabe}
\author[r]{and N.~Yokozaki}

\affiliation[a]{Dipartimento di Fisica e Astronomia ``Galileo Galilei'', Universit\`a di Padova, Via Marzolo 8, I-35131 Padova, Italy}
\affiliation[b]{INFN, Sezione di Padova, Via Marzolo 8, I-35131 Padova, Italy}
\affiliation[c]{Centre de Physique Th\'eorique, CNRS, \'Ecole Polytechnique, IP Paris, 91128 Palaiseau, France}
\affiliation[d]{Laboratoire de Physique Th\'eorique (UMR8627), CNRS, Univ. Paris-Sud, Universit\'e Paris-Saclay, 91405 Orsay, France}
\affiliation[e]{INFN,  Sezione di Roma Tre, Via della Vasca Navale 84, I-00146 Roma, Italy}
\affiliation[f]{Paul Scherrer Institut, CH-5232 Villigen PSI, Switzerland}
\affiliation[g]{IFIC, Universitat de Val\`{e}ncia - CSIC, Apt.~Correus 22085, E-46071, Val\`encia, Spain}
\affiliation[h]{Dept.~de F\'{\i}sica Qu\`antica i Astrof\'{\i}sica, Institut de Ci\`encies del Cosmos (ICCUB), Universitat de Barcelona, Mart\'i Franqu\`es 1, E-08028 Barcelona, Spain}
\affiliation[i]{INFN, Sezione di Roma, Piazzale A. Moro 2, I-00185 Roma, Italy}
\affiliation[j]{Institute of Theoretical Physics, Faculty of Physics, University of Warsaw, ul. Pasteura 5, PL–02–093 Warsaw, Poland}
\affiliation[k]{Theory Center, IPNS, KEK, Tsukuba 305-0801, Japan}
\affiliation[l]{DESY, Notkestra{\ss}e 85, D-22607 Hamburg, Germany}
\affiliation[m]{Institut f\"ur Physik, Humboldt-Universit\"at zu Berlin, D-12489 Berlin, Germany}
\affiliation[n]{CERN, Geneva, Switzerland}
\affiliation[o]{Physics Department, Florida State University, Tallahassee, FL 32306-4350, USA}
\affiliation[p]{Theoretical Physics Department, CERN, Geneva, Switzerland}
\affiliation[q]{Department of Physics and Astronomy, University of California, Irvine, CA 92697 USA}
\affiliation[r]{Department of Physics, Tohoku University, Sendai, Miyagi 980-8578, Japan}

\abstract{\HEPfit is a flexible open-source
tool which, given the Standard Model or any of its extensions, allows to \textit{i)} fit the model
parameters to a given set of experimental observables;
\textit{ii)} obtain predictions for observables.
\HEPfit can be used either in Monte Carlo mode, to perform a Bayesian Markov Chain Monte Carlo
analysis of a given model, or as a library, to obtain predictions of
observables for a given point in the parameter space of the model, allowing \HEPfit to be used in
any statistical framework. In the present version, around a thousand observables have been implemented
in the Standard Model and in several new physics scenarios. In this paper, we describe the general structure of the code as well as models and observables implemented in the current release.}

\emailAdd{hepfit-support@roma1.infn.it}

\begin{document}

\begin{center}
\begin{tabular*}{\textwidth}{p{8cm} r r}
         \multirow{5}{*}{\href{https://hepfit.roma1.infn.it/}{\includegraphics[width=0.2\textwidth]{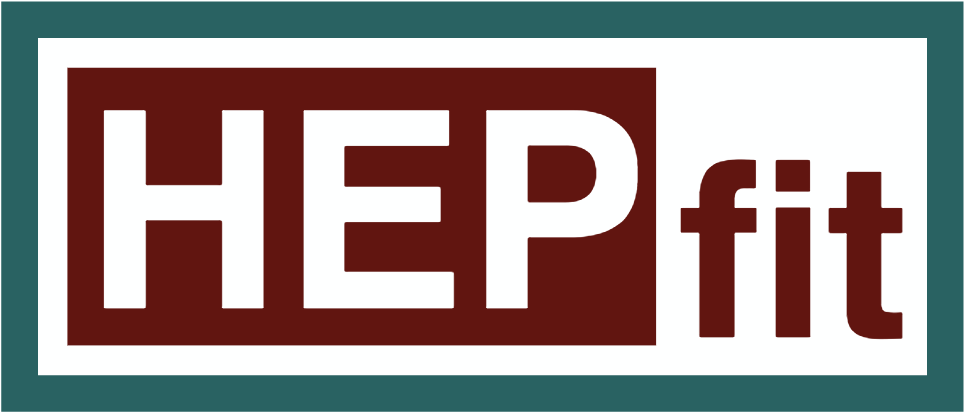}}}\hfill
         & CERN-TH-2019-178
         & CPHT-RR060.102019 \\
         & DESY 19-184
         & FTUV/19-1031 \\
         & HU-EP-19/34
         & IFIC/19-44 \\
         & KEK-TH-2163 
         & LPT-Orsay-19-36 \\
         & PSI-PR-19-22 
         & UCI-TR-2019-26
\end{tabular*}
\end{center}

\maketitle

%%%%%%%%%%%%%%%%%%%%%%%%%%%%%%%%%%%%%
\section{Introduction}
%%%%%%%%%%%%%%%%%%%%%%%%%%%%%%%%%%%%%

Searching for New Physics (NP) beyond the Standard Model (SM) in the era of the Large
Hadron Collider (LHC) requires combining
experimental and theoretical information from many sources to optimize
the NP sensitivity. NP searches, even in the absence of a positive
signal, provide useful information which puts constraints on the
viable parameter space of any NP model. Should a NP signal emerge at
future LHC runs or elsewhere, the combination of all available
information remains a crucial step to pin down the actual NP model.
NP searches at the LHC require extensive detector simulations and are
usually restricted to a subset of simplified NP models. Given the high
computational demand of direct searches, it is crucial to explore only
regions of the parameter space compatible with other constraints. In
this respect, indirect searches can be helpful and make the study of
more general models viable.

\HEPfit aims at providing a tool which allows to combine all
available information to select allowed regions in the parameter space
of any NP model. To this end, it can compute many observables with
state-of-the-art theoretical expressions in a set of models which can
be extended by the user. It also offers the possibility of sampling
the parameter space using a Markov Chain Monte Carlo (MCMC) implemented using
the \BAT library~\cite{2009CoPhC.180.2197C,Caldwell_2010,Beaujean_2011}. Alternatively, \HEPfit can be
used as a library to obtain predictions of the observables in any
implemented model. This allows to use \HEPfit in any statistical
framework.

\HEPfit is written in \texttt{C++} and parallelized with MPI. This is the first public release with a limited set of observables and
models, which we plan to enlarge. The code is released under the GNU
General Public License, so that contributions from users are possible
and welcome. In particular, the present version provides Electroweak
Precision Observables (EWPO), Higgs signal strengths, and several
flavour observables in the SM, in Two-Higgs-Doublet
Models (THDM), and in several parameterizations of NP contributions. Furthermore, it also calculates various Lepton Flavour Violating (LFV) observables in the Minimal Supersymmetric Standard Model (MSSM). In the near future, we plan to add many more flavour observables and to enlarge
the spectrum of NP models.

The paper is organized as follows. In Section~\ref{sec:Code} we give a brief description of \HEPfit including the statistical framework used, the MPI parallelization and some other details. In Section~\ref{sec:Models} we discuss the models implemented in \HEPfit. In Section~\ref{sec:Observables} we go on discussing some of the observables implemented in \HEPfit. In Section~\ref{sec:Results} we present some physics results obtained using \HEPfit in previous publications. Indeed several physics analyses~\cite{Ciuchini:2013pca,Reina:2015yuh,Ciuchini:2015qxb,deBlas:2016ojx,Cacchio:2016qyh,Ciuchini:2016weo,deBlas:2016nqo,Ciuchini:2017mik,Ciuchini:2017gva,Eberhardt:2017ulj,Gori:2017tvg,deBlas:2017wmn,Chowdhury:2017aav,deBlas:2018tjm,Chiang:2018cgb,Cheng:2018mkc,Eberhardt:2018lub,Ciuchini:2018anp,Ciuchini:2018xll,Buccella:2019kpn,Ciuchini:2019usw,Arnan:2019uhr,Becirevic:2019tpx,deBlas:2019rxi,deBlas:2019wgy,Durieux:2019rbz,Darme:2020hpo} have been completed using HEPfit and serves as a validation of
the code and its use as an open-source computational framework. A detailed description of the installation procedure can be found in Section~\ref{sec:Installation} followed by examples of how to use \HEPfit in Section~\ref{sec:Usage}. Updated information and detailed online documentation can be found on
the \href{https://hepfit.roma1.infn.it/}{\HEPfit website} \cite{website}.

%%%%%%%%%%%%%%%%%%%%%%%%%%%%%%%%%%%%%
\section{The \HEPfit code}
\label{sec:Code}
%%%%%%%%%%%%%%%%%%%%%%%%%%%%%%%%%%%%%
\HEPfit is a computational tool for the combination of indirect and
direct constraints on High Energy Physics models. The code is built in
a modular structure so that one can pick and choose which observables
to use and what model to analyze. It also provides an interface that
can be used to build customized models along with customized
observables. This flexible framework allows defining a model of choice and observables that depend on the parameters of this model, thus opening up the possibility of using \HEPfit to perform phenomenological analyses in such a model.

The tool comes with a built-in statistical framework
based on a Bayesian MCMC analysis.
However, any statistical framework can be used along with this tool
since a library is made available. \HEPfit also allows for the incorporation of parametric and experimental correlations and can read likelihood distributions directly from \ROOT histograms. This removes the necessity for setting experimental constraints through parameterized distributions which might require making approximations.

Since the statistical core of \HEPfit is based on a MCMC, speed of
computation is of utmost importance. \HEPfit is already massively
parallelized to run over large number of CPUs using \texttt{OpenMPI}
and scales well to hundreds of processing units. The framework further brings forth
the flexibility of defining a model of choice and observables that
depend on the parameters of this model, thus opening up the possibility
of performing various analyses using \HEPfit. 

The package comes with
several examples of how \HEPfit can be used and detailed documentation
of the code and the physics can be found online on the 
\href{https://hepfit.roma1.infn.it/}{\HEPfit website}. Throughout
its development, emphasis has been placed on speed and streamlining
error handling. \HEPfit has been tested through several analyses on various
hardware architecture and displays reliable scaling to large systems.

%%%%%%%%%%%%%%%%%%%%%%%%%%%%%%%%%%%%%
\subsection{Statistical framework}
\label{sec:Stat}
%%%%%%%%%%%%%%%%%%%%%%%%%%%%%%%%%%%%%

\HEPfit can be used both as a library to compute the values of chosen
observables and also as a MCMC based Bayesian
analysis framework. While the former approach allows for choosing the
statistical framework one wants to use, the latter uses a robust
Bayesian MCMC framework implemented in the public code
\BAT~\cite{2009CoPhC.180.2197C,Caldwell_2010,Beaujean_2011}. In this
section we give a brief overview of the Bayesian statistical framework implemented in \HEPfit using \BAT.

%%%%%%%%%%%%%%%%%%%%%%%%%%%%%%%%%%%%%
\subsubsection{Bayesian framework}
\label{sec:Bayes}
%%%%%%%%%%%%%%%%%%%%%%%%%%%%%%%%%%%%%

Once the model parameters, $\vec{\mathbf{x}}$, and the data, $D$, are defined one can define the posterior distribution according to Bayes theorem as:

\begin{equation}
    P(\vec{\mathbf{x}}|D)=\frac{P(D|\vec{\mathbf{x}})P_0(\vec{\mathbf{x}})}{\int P(D|\vec{\mathbf{x}})P_0(\vec{\mathbf{x}})d\vec{\mathbf{x}}}~,
    \label{eq:Bayes}
\end{equation}
where $P_0(\vec{\mathbf{x}})$ is the prior distribution of the
parameters which represents the prior knowledge of the parameters
which can come from experiments or theory computations or can be
uninformative. The denominator is called the {\em normalization} or the
{\em evidence}, the computation of which can allow for model comparison
through the Bayes factor. The likelihood is denoted as
$P(D|\vec{\mathbf{x}})$. Once the (unnormalized) posterior
distribution\footnote{While for a simple parameter space it is possible to compute the normalization factor, a MCMC analysis
  provides an unnormalized posterior distribution. This is the relevant ingredient for the purpose of studying any credibility
  interval. From now on, we implicitly assume we are dealing with an
  unnormalized posterior density.} is mapped out using sampling
methods (in our case a MCMC routine), one can obtain the marginalized
posterior distributions of the individual parameters from which the
credibility regions can be computed. The 1D marginalized distribution
is given by
\begin{equation}
    P(x_i|D) = \int P(\vec{\mathbf{x}}|D) \prod_{j \ne i} d x_{j}~,
    \label{eq:margin}
\end{equation}
where all the variables but the one for which the marginalized
posterior distribution is being computed are integrated over, and
similarly for marginalized 2D distributions.

%%%%%%%%%%%%%%%%%%%%%%%%%%%%%%%%%%%%%
\subsubsection{Markov Chain Monte Carlo}
\label{sec:MCMC}
%%%%%%%%%%%%%%%%%%%%%%%%%%%%%%%%%%%%%
In general, the posterior distribution specified in
Eq.~(\ref{eq:Bayes}) cannot be computed easily, especially when there is a proliferation of model parameters. 
% The same applies for the marginalized distributions. 
Using a naive Monte Carlo sampling algorithm can lead to unacceptable execution times because of their inherent inefficiency in sampling the parameter space. However, MCMC procedures overcome this hurdle and make the
application of Bayes theorem quite tractable.

The implementation of MCMC in \BAT uses a Metropolis-Hastings
algorithm to sample the parameter space from the posterior. The steps of a Metropolis-Hastings algorithm for sampling from a (unnormalized) probability
density $f(\vec{\mathbf{x}})$ are as follows:
\begin{enumerate}
    \item Start at a random point in the parameter space $\vec{\mathbf{x}}$.
    \item Generate a proposal point $\vec{\mathbf{y}}$ according to a
      symmetric probability distribution $g(\vec{\mathbf{x}},\vec{\mathbf{y}})$.
    \item Compare the value of the function $f$ at proposal point
      $\vec{\mathbf{y}}$ with the value at the current point $\vec{\mathbf{x}}$. 
    The proposal point is accepted if:
    \begin{itemize}
        \item $f(\vec{\mathbf{y}}) \ge f(\vec{\mathbf{x}})$,
        \item otherwise, generate a random number $r$ from a uniform distribution in the range $[0,1]$ and accept the proposal if $f(\vec{\mathbf{y}})/f(\vec{\mathbf{x}}) > r$.
    \end{itemize}
    If neither conditions are satisfied the proposal is rejected.
  \item Continue from step 1.
\end{enumerate}
In our case, the function $f(\vec{\mathbf{y}})$ is the unnormalized 
posterior, namely the numerator of Eq.~(\ref{eq:Bayes}). 

The MCMC implementation consists of two parts. The first part
is called the pre-run or the burn-in phase where the chains start from
arbitrary random points in the parameter space and reach a stationary
state after a certain number of iterations, through the tuning of the
proposal function. The stationary state is reached once the targeted
efficiency of the proposal and $R$-values close to one are obtained. The $R$-value
for a parameter is essentially the distance of its mean values in the
various chains in units of the standard deviation of the parameter in
each chain \cite{10.2307/2246093,doi:10.1080/10618600.1998.10474787}. Samples of
the parameter space are not collected during the pre-run. Once the
pre-run is over, the samples of the parameter space are collected in
the main run to get the marginalized distributions of all the
parameters and the corresponding posterior distributions of the
observables and of any other derived quantity that may have been
defined. The details of the implementation of the MCMC framework can
be found in
Refs.~\cite{2009CoPhC.180.2197C,Caldwell_2010,Beaujean_2011}.

In our work we have not faced any limitations to the number of parameters that can be used and the number of observables that can be used in the fit. However, one has to take note of the following regarding the time taken for the fits:
\begin{itemize}
    \item The convergence of the Markov chains is slower for larger number of parameters and when the parameters are correlated explicitly or as a consequence of the data used as observables for both factorized and non-factorized priors. To reduce the time of the fit it is best to reduce the number of parameters to the minimum necessary.
    \item For larger number of parameters ($>$ 30) it is advised to compare fits using both the factorized and non-factorized priors for optimal performance (see Section~\ref{sec:Usage} for details).
    \item There is no limitation to the number of observables that can be defined. However, if the observables are computationally expensive, they will slow the fits accordingly.
\end{itemize}

The largest fits that we have performed with \HEPfit contained more than 90 parameters fit to more than 200 observables. Other fits have been done with several hundred observables but smaller number of parameters. We have not seen any limitation from these other than the time consumed to do the fits, which are at most a few days as shown in Table~\ref{tab:MPI}.

%%%%%%%%%%%%%%%%%%%%%%%%%%%%%%%%%%%%%
\subsubsection{Integration of \BAT with \HEPfit}
\label{sec:overloading}
%%%%%%%%%%%%%%%%%%%%%%%%%%%%%%%%%%%%%
The MCMC framework implemented in \BAT is integrated in \HEPfit using the library that is provided by \BAT on compilation. The \texttt{MonteCarloEngine} class in \HEPfit inherits from the \texttt{BCModel} class in \BAT and overloads the \texttt{LogLikelihood} function. This method generates the numerical likelihood for one point in the parameter space with the values of the observables computed by \HEPfit and the experimental and theoretical constraints provided to \HEPfit. The parameters and their distributions are passed by \HEPfit to \BAT through the \texttt{MonteCarloEngine} class. \HEPfit takes care of correlated parameter priors by rotating them to the eigenvector basis in order
to increase the efficiency of sampling. 

The output of a run, as detailed in Section~\ref{sec:MC}, is produced
by both \BAT and \HEPfit. All 1D and 2D marginalized distributions and
posterior distributions are produced using the \texttt{BCH1D} and the
\texttt{BCH2D} classes of BAT and stored in a \ROOT file. One
can choose to store the chains in the \ROOT file as well using
\HEPfit. While this is useful for post-processing, since it makes the
full sample available point by point, it entails a dramatic increase
in size of the output \ROOT file. 

It should be noted that \BAT is necessary only when running in the
MCMC mode. If one chooses to run \HEPfit as an event generator or to
only compute values of observables for custom statistical analyses,
the interface with \BAT is not used at all.

%%%%%%%%%%%%%%%%%%%%%%%%%%%%%%%%%%%%%
\subsection{Parallelization with MPI}
\label{sec:MPI}
%%%%%%%%%%%%%%%%%%%%%%%%%%%%%%%%%%%%%
{\footnotesize
\begin{table}[t!]
    \centering
    \begin{tabular}[]{|p{4.8cm}|l|l|l|}
        \hline
        {\bf Physics Problem} &  {\bf Hardware } & {\bf Run Configuration} & {\bf Time} \\
        \hline
        \multirow{2}{*}{Unitarity Triangle Fit}& 3 nodes, 120 CPUs& 120 chains, 1.4M iterations& 00:02:10\\
                                               & 1 nodes, 40  CPUs& 40 chains, 600K iterations& 00:00:21\\
        \hline                                       
        \multirow{2}{4.8cm}{$b \to s$ decays in SMEFT$^\dagger$ \cite{Ciuchini:2019usw}}    & 6 nodes, 240 CPUs& 240 chains, 12.5K iterations & 02:05:00\\
                                               & 6 nodes, 240 CPUs& 240 chains, 39K iterations& 05:20:00\\
        \hline
        \multirow{2}{4.8cm}{combination of 
        Higgs signal strengths and EWPO\cite{deBlas:2019wgy}}       &1 node, 16 CPUs& 16 chains, 5M iterations& 00:14:15 \\
                                               &1 node, 16 CPUs& 16 chains, 24M iterations& 02:08:00 \\
        \hline
        \multirow{2}{4.8cm}{$D\to PP$ decays and CP asymmetry\cite{Buccella:2019kpn}}      &3 nodes, 240 CPUs& 240 chains, 4M iterations& 00:18:30\\
                                                                    & 1 node, 8 CPUs& 8 chains, 200K iterations& 00:00:10\\
        \hline
    \end{tabular}
    \caption{{\it Some representative runs with} \HEPfit ~{\it to show the advantages of the MPI implementation. Times are given in DD:HH:MM. The number of iterations refer to the sum total of pre-run and main-run iterations. The number of chains are equal to the number of CPUs by choice. $^\dagger$The $b\to s$ analysis is done with factorized priors, hence the number of iterations should be multiplied by the number of parameters ($\sim$50) to get a comparative estimate with the other cases. All runs performed in the BIRD or Maxwell clusters at DESY, Hamburg.}}
    \label{tab:MPI}
\end{table}
}
One of the most important advantages of \HEPfit over several other
similar publicly available codes is that it is completely parallelized
using \texttt{OpenMPI}, allowing it to be run on both single CPUs with multiple
cores and on several nodes on large clusters. The MCMC algorithm is
very apt for this kind of parallelization since an integer number of
chains can be run on each core. Ideally allocating one core
per chain minimizes the run time.

The official version of \BAT is parallelized using \texttt{OpenMP}. However,
\texttt{OpenMP} relies on shared memory and cannot be distributed over several
nodes in a cluster. To overcome this limitation we used \texttt{OpenMPI} to
parallelize both \BAT and \HEPfit. The parallelization is at the level
of the computation of likelihood and observables. 
This means that the MCMC at
both the pre-run and main run stages can take advantage of this
parallelization. Once the likelihood computation (which requires the
computation of the observables) is done, the flow is returned to the
master, which performs the generation of the next set of proposal
points. The computation of efficiencies and convergence, as well as the
optimization of the proposal function, are currently not distributed
since they require full information on the chain states. This is the
only bottleneck in the parallelization, since the time the master
takes to process these steps might be comparable to the time
required to compute all the observables by each chain, if the number
of chains is very large. However, this begins to be a matter of
concern only when the number of chains is in the range of several
hundreds, a situation that a normal user is unlikely to encounter.

To demonstrate the advantages that one can get from the
parallelization built into \HEPfit and to give an estimate of the
scaling of the run-times with the number of cores, we give some
examples of analyses that can be done both on personal computers and
on large clusters in Table~\ref{tab:MPI}. These should not be taken as benchmarks since we do
not go into the details of the hardware, compiler optimization,
etc. Rather, these should be taken as an indication of how MPI
parallelization greatly enhances the performance of the \HEPfit code.

%%%%%%%%%%%%%%%%%%%%%%%%%%%%%%%%%%%%%
\subsection{Custom models and observables}
\label{sec:mymodel}
%%%%%%%%%%%%%%%%%%%%%%%%%%%%%%%%%%%%%

Another unique feature that \HEPfit offers is the possibility of
creating custom models and custom observables. All the features of
\HEPfit are made available along with all the observables and
parameters predefined in \HEPfit. An example of such a use of \HEPfit
can be found in Ref.~\cite{Buccella:2019kpn}. Detailed instructions
for implementation are given in Section~\ref{sec:custom}.

The user can define a custom model using a template provided with the
package, by adding a set of parameters to any model defined in
\HEPfit. Generally, in addition to defining the new parameters, the
user should also specify model-specific additional contributions to
any observables predefined in \HEPfit that he wants to
use. Furthermore, new observables can be defined in terms of these new
parameters.

New observables can also be defined in the context of the predefined
\HEPfit models. In this case, the user just needs to specify the
observable in terms of the model parameters, without the need to
create a custom model. The parameters already used in \HEPfit
can also be accessed. For example, one does not need to redefine the
Cabibbo-Kobayashi-Maskawa (CKM) mixing matrix, $V_{\mathrm{CKM}}$,
%mass mixing matrix, $V_{\mathrm{CKM}}$, 
if one needs to use it in the
computation of a custom observable. One can simply call the SM object
available to all observables and then use the implementation of
$V_{\mathrm{CKM}}$ already provided either in terms of the Wolfenstein
parameters or in terms of the elements of the matrix. It should be
noted that one does not need to define a custom model to define custom
observables. A custom model should be defined only if the user
requires parameters not already present in \HEPfit. More details can
be found in Section~\ref{sec:custom}.

%%%%%%%%%%%%%%%%%%%%%%%%%%%%%%%%%%%%%
\section{Models defined in \HEPfit}
\label{sec:Models}
%%%%%%%%%%%%%%%%%%%%%%%%%%%%%%%%%%%%%

The basic building blocks of \HEPfit are the classes \texttt{Model}
and \texttt{Observable}.  Actual models extend the base class
\texttt{Model} sequentially (e.g. \texttt{QCD} $\leftarrow$
\texttt{StandardModel} $\leftarrow$ \texttt{THDM} $\leftarrow$
\dots). Inheritance allows a given model to use all the methods of the
parent ones and to redefine those which have to include
additional contributions specific to the extended model.  For example, the method
computing the strong coupling constant ($\alpha_s$) includes strong
corrections in \texttt{QCD}, adds electromagnetic corrections in
\texttt{StandardModel}, and any
additional contributions in classes extending the \texttt{StandardModel}.  Models contain
model parameters (both fundamental model parameters and auxiliary
ones) and model flags which control specific options.

An instance of the \texttt{Observable} class contains the experimental
information relative to a given physical observable as well as an
instance of the class \texttt{ThObservable}, responsible for the
computation of that observable in the given model. This is the class
where both the experimental or theoretical constraints and the theory
computation in the model are accessible, allowing for the likelihood calculation. We now briefly review the models implemented in the current release of \HEPfit.

%%%%%%%%%%%%%%%%%%%%%%%%
\subsection{The Standard Model}
\label{sec:SM}
%%%%%%%%%%%%%%%%%%%%%%%%

In \HEPfit, the minimal model to be defined in order to compute any
observable is the 
\texttt{StandardModel}, which for convenience extends a class
\texttt{QCD}, which in turn, extends the abstract class
\texttt{Model}.

The model implemented in the \texttt{QCD} class defines the following
model parameters: the value of $\alpha_s(M)$ at a provided scale $M$, the
$\overline{\mathrm{MS}}$ quark masses ${\bar m}_q$ (except
for the top quark mass where for convenience the pole mass is taken as
input parameter and then converted to $\overline{\mathrm{MS}}$). With
this information, the class initializes instances of the
$\texttt{Particle}$ class for each quark. In addition, objects of type
$\texttt{Meson}$, containing information on masses, lifetimes, decay
constants and other hadronic parameters (these are taken as model
parameters although in principle they are derived quantities), are
instantiated for several mesons. Furthermore, bag parameters for meson
mixings and decays are instantiated. This class also defines methods to
implement the running of $\alpha_s$ and quark masses.

The \texttt{StandardModel} class extends \texttt{QCD} by adding the
remaining SM parameters, namely the Fermi constant $G_F$, the
fine-structure constant $\alpha$, the $Z$ boson mass $M_Z$, the Higgs
boson mass $m_h$ and the CKM mixing matrix
%Cabibbo-Kobayashi-Maskawa (CKM) mixing matrix
(instantiating the corresponding object \texttt{CKM}).\footnote{Two
  CKM parameterizations (the Wolfenstein one and a parameterization
  using $\vert V_{us}\vert$, $\vert V_{cb}\vert$, $\vert V_{ub}\vert$,
  and the angle $\gamma$ of the unitarity triangle as inputs) can be
  selected using the model flag \texttt{FlagWolfenstein}.} The
Pontecorvo-Maki-Nakagawa-Sakata (PMNS) mixing matrix is defined but
currently not activated. It also fixes the \texttt{QCD} parameter $M$
introduced above to $M_{Z}$ and the \texttt{QCD} parameter
$\alpha_{s}(M)$ to $\alpha_{s}(M_{Z})$.  Furthermore, it contains
$\texttt{Particle}$ objects for leptons.  Several additional model
parameters describe the hadronic vacuum polarization contribution to
the running of $\alpha$, and the theoretical uncertainties in the $W$
mass and other EWPO, for which is convenient to
use available numerical estimates. Moreover, the running of
$\alpha_s$ is extended to include electromagnetic corrections.

The \texttt{StandardModel} class also provides matching conditions for weak effective Hamiltonians through the class \texttt{StandardModelMatching}. Low-energy weak effective Hamiltonians, both $\Delta F=1$ and $\Delta F=2$, are provided on demand by the class \texttt{Flavour} instantiated by \texttt{StandardModel}.

Although extending \texttt{Model} and \texttt{QCD}, \texttt{StandardModel} is the actual base class
for any further definition of NP models (e.g. THDM, SUSY, etc.). Details on the implementation of \texttt{StandardModel} and \texttt{QCD} can be found in the \href{https://hepfit.roma1.infn.it/doc/latest-release/index.html}{online documentation}.

%%%%%%%%%%%%%%%%%%%%%%%%
\subsection{Two-Higgs-Doublet models}
\label{sec:THDM}
%%%%%%%%%%%%%%%%%%%%%%%%

One of the most straightforward extensions of the SM is
the Two-Higgs-Doublet model (THDM)
\cite{Lee:1973iz,Gunion:2002zf,Branco:2011iw}. No fundamental theorem
forbids to add a second scalar doublet to the SM particle
content. The THDM can offer a solution to problems as the stability of
the scalar potential up to very large scales (see
e.g. ref.~\cite{Chowdhury:2015yja}) or electroweak baryogenesis (see
e.g. refs.~\cite{Bochkarev:1990fx,Nelson:1991ab,Dorsch:2013wja}),
which cannot be solved in the SM. Furthermore, it could
emerge as an effective description of more complicated models like
SUSY models, which necessarily contain two Higgs
doublets. 

There are several THDM variants with different phenomenological
implications. At the moment \HEPfit contains the versions which
exclude flavour-changing neutral currents at tree-level as well as CP
violation in the Higgs sector. In order to fulfil the first demand, an
additional softly broken $Z_2$ symmetry is assumed, which can be
chosen in four different ways; thus these versions are called type I,
type II, type X and type Y.\footnote{The THDM of type II contains the scalar Higgs part of the MSSM.} The four types only differ in the Yukawa couplings of the Higgs fields. The corresponding assignments can be found in Table \ref{tab:THDMtypes}, where $Y^f_j$ denotes the coupling
of one of the two Higgs doublets $\Phi_j\ (j=1,2)$ to the fermion field
$f$.
\begin{table}[tb]
  \centering
  \begin{tabular}{|l|l|l|l|}
    \hline
      Type I & Type II & Type X (``lepton specific'') & Type Y (``flipped'') \\
    \hline
      $Y^d_{1}\equiv 0$, $Y^\ell_{1}\equiv 0$ & $Y^d_{2}\equiv 0$, $Y^\ell_{2}\equiv 0$ & $Y^d_{1}\equiv 0$, $Y^\ell_{2}\equiv 0$ & $Y^d_{2}\equiv 0$, $Y^\ell_{1}\equiv 0$ \\
      \hline
  \end{tabular}
  \caption{Yukawa couplings in the four possible $Z_2$ symmetric THDM types.}
 \label{tab:THDMtypes}
\end{table} 
By definition, $Y_1^u \equiv 0$ for all four types. In the configuration file \texttt{THDM.conf}, one has to choose the THDM type by setting the flag \texttt{modelTypeflag} to \texttt{type1}, \texttt{type2}, \texttt{typeX} or \texttt{typeY}.

We write the Higgs potential for $\Phi_1$ and $\Phi_2$ as
\begin{align}
V_H^\text{\tiny{THDM}} & = m_{11}^2\Phi_1^\dagger\Phi_1^{\phantom{\dagger}}
 	+m_{22}^2\Phi_2^\dagger\Phi_2^{\phantom{\dagger}}
 	-m_{12}^2\left[ \Phi_1^\dagger\Phi_2^{\phantom{\dagger}}
 		+\Phi_2^\dagger\Phi_1^{\phantom{\dagger}}\right] 
	+\tfrac12 \lambda_1(\Phi_1^\dagger\Phi_1^{\phantom{\dagger}})^2 
	+\tfrac12 \lambda_2(\Phi_2^\dagger\Phi_2^{\phantom{\dagger}})^2 \nonumber \\
&\phantom{{}={}}
 	+\lambda_3(\Phi_1^\dagger\Phi_1^{\phantom{\dagger}})
 		(\Phi_2^\dagger\Phi_2^{\phantom{\dagger}})
 	+\lambda_4(\Phi_1^\dagger\Phi_2^{\phantom{\dagger}})
 		(\Phi_2^\dagger\Phi_1^{\phantom{\dagger}})
	+\tfrac12 \lambda_5^{} \left[ (\Phi_1^\dagger\Phi_2^{\phantom{\dagger}})^2
 		+(\Phi_2^\dagger\Phi_1^{\phantom{\dagger}})^2 \right], \label{eq:THDMVH}
\end{align}
and the Yukawa part of the Lagrangian as
 \begin{align}
{\cal L}_Y^\text{\tiny{THDM}} &= -Y^u_2 \bar Q_{\textit{\tiny{L}}} \tilde\Phi_2 u_{\textit{\tiny{R}}} -\sum\limits_{j=1}^2 \left[ Y^d_j \bar Q_{\textit{\tiny{L}}} \Phi_j d_{\textit{\tiny{R}}} +Y^\ell_j \bar L_{\textit{\tiny{L}}} \Phi_j \ell_{\textit{\tiny{R}}}\right] + {\text{h.c.}} \nonumber ,
\end{align}
 where one of the choices from Table \ref{tab:THDMtypes} has to be applied.

The THDM contains five physical Higgs bosons, two of which are neutral and even under CP transformations, one is neutral and CP-odd, and the remaining two carry the electric charge $\pm$1 and are degenerate in mass. We assume that the 125 GeV resonance measured at the LHC is the lighter CP-even Higgs $h$, while the other particles are labelled $H$, $A$ and $H^\pm$, respectively.
The eight parameters from the Higgs potential \eqref{eq:THDMVH} can be transformed into physical parameters:
\begin{itemize}
\item the vacuum expectation value $v$,
\item the lighter CP-even Higgs-boson mass $m_h$,
\item the heavier CP-even Higgs-boson mass $m_H$,
\item the CP-odd Higgs-boson mass $m_A$,
\item the charged Higgs-boson mass $m_{H^+}$,
\item the mixing angle $\alpha$,
\item the mixing angle $\beta$ and
\item the soft $Z_2$ breaking parameter $m_{12}^2$ from \eqref{eq:THDMVH}.
\end{itemize}
The Fermi constant $G_F$ and $m_h$ are defined in the SM configuration file. For practical reasons, the \HEPfit implementation uses $\beta-\alpha$ and $\log_{10}\tan\beta$, instead of $\alpha$ and $\beta$, and squared $H$, $A$ and $H^+$ masses. 

%%%%%%%%%%%%%%%%%%%%%%%%
\subsection{The Georgi-Machacek model}
\label{sec:GMmodel}
%%%%%%%%%%%%%%%%%%%%%%%%

In the Georgi-Machacek model \cite{Georgi:1985nv,Chanowitz:1985ug}, the SM is extended by two $SU(2)$ triplets. This construction can simultaneously explain the smallness of neutrino masses (via the seesaw mechanism) and the electroweak $\rho$ parameter. In \HEPfit, we implemented the custodial Georgi-Machacek model, in which the additional heavy scalars can be combined into a quintet, a triplet and a singlet under the custodial $SU(2)$ with masses $m_5$, $m_3$, and $m_1$, respectively. Further model parameters in \HEPfit are the triplet vev $v_\Delta$, the singlet mixing angle $\alpha$ and the two trilinear couplings $\mu_1$ and $\mu_2$. For details of the \HEPfit implementation of this model we refer to reference \cite{Chiang:2018cgb}.

%%%%%%%%%%%%%%%%%%%%%%%%
\subsection{Oblique corrections in electroweak precision observables }
\label{sec:Oblique}
%%%%%%%%%%%%%%%%%%%%%%%%

Assuming the physics modifying the on-shell properties of the $W$ and $Z$ bosons is universal, 
%all effects in the so-called EWPO ($Z$-pole data and $W$ mass and decay widths)
such effects
can be encoded in three quantities: 
the relative normalization of neutral and charged currents, and the two relative differences between the
three possible definitions of the weak mixing angle. These effects are captured by the so-called $\epsilon_i$ 
parameters introduced in \cite{Altarelli:1990zd,Altarelli:1991fk,Altarelli:1993sz}. 
The model class {\tt NPEpsilons\_pureNP} describes the NP contributions to these quantities. 
It also allows contributions in the additional, non-universal, $\epsilon_b$ parameter, also introduced 
in \cite{Altarelli:1993sz} to describe modifications of the $Zb\bar{b}$ interactions. The model parameters 
in this class are defined in Table~\ref{tab:Epsilons}.

 \begin{table}[tb]
 \centering
  \begin{tabular}{| c c | }
\hline
\textbf{\HEPfit name}&
\textbf{Parameter}\\
\hline
 {\tt delEps\_1}&
$\delta \epsilon_1$\\
{\tt delEps\_2}&
$\delta \epsilon_2$\\
{\tt delEps\_3}&
$\delta \epsilon_3$\\
{\tt delEps\_b}&
$\delta \epsilon_b$\\
\hline
  \end{tabular}
  \caption{Model parameters in the {\tt NPEpsilons\_pureNP} class.}
 \label{tab:Epsilons}
\end{table} 
The $\delta \epsilon_i$, $i=1,2,3$ can be readily mapped into the oblique parameters describing NP 
modifying the propagator of the electroweak gauge bosons:
\begin{eqnarray}
\delta \varepsilon_1&=&\alpha T-W + 2 X \frac{\sin{\theta_w}}{\cos{\theta_w}}-Y \frac{\sin^2{\theta_w}}{\cos^2{\theta_w}},\\
\delta \varepsilon_2&=&-\frac{\alpha}{4 \sin^2{\theta_w}}U-W + 2 X \frac{\sin{\theta_w}}{\cos{\theta_w}}-V,\\
\delta \varepsilon_3&=&\frac{\alpha}{4 \sin^2{\theta_w}}S-W +  \frac{X}{\sin{\theta_w}\cos{\theta_w}}-Y,
\end{eqnarray}
where $\theta_w$ is the weak mixing angle, the $S,T,U$ parameters were originally introduced in Ref. \cite{Peskin:1991sw} and $V,W,X,Y$ in Ref.~\cite{Barbieri:2004qk}. 
All these parameterize the different coefficients in the expansion of the gauge boson self-energies 
for $q^2\ll \Lambda^2$ with $\Lambda$ the typical scale of the NP. Traditionally, the literature of 
electroweak precision tests has focused on the first three parameters (which also match the number of different 
universal effects that can appear in the EWPO). Because of that, we
include the model class {\tt NPSTU}, which describes this type of NP. The relevant parameters are 
collected in Table~\ref{tab:STU}. 
It is important to note, however, that the $U$ parameter is typically expected 
to be suppressed with respect to $S,T$ by $M_W^2/\Lambda^2$. Indeed, at the leading order in 
$M_W^2/\Lambda^2$ the four parameters describing universal NP effects in electroweak observables are $S,T,W$ and $Y$~\cite{Barbieri:2004qk}.
%The adequate dependence is properly parameterized by the $\epsilon_i$ parameters above. 
%
 \begin{table}[tb]
 \centering
  \begin{tabular}{| c c | }
\hline
\textbf{\HEPfit name}&
\textbf{Parameter}\\
\hline
 {\tt obliqueS}&
$S$\\
{\tt obliqueT}&
$T$\\
{\tt obliqueU}&
$U$\\
\hline
  \end{tabular}
  \caption{Model parameters in the {\tt NPSTU} class.}
 \label{tab:STU}
\end{table} 

%%%%%%%%%%%%%%%%%%%%%%%%
\subsection{The dimension-six Standard Model Effective Field Theory }
\label{sec:Dim6}
%%%%%%%%%%%%%%%%%%%%%%%%

When the typical mass scale of NP is significantly larger than the energies tested by the experimental 
observables, the new effects can be described in a general way by means of an effective Lagrangian
\begin{equation}
{\cal L}_{\mathrm{eff}}={\cal L}_{\mathrm{SM}} + \sum_{d>4} \frac{1}{\Lambda^{d-4}}{\cal L}_d.
\label{eq:EffLag}
\end{equation}
In Eq.~(\ref{eq:EffLag}) ${\cal L}_{\mathrm{SM}}$ is the SM Lagrangian, $\Lambda$ is the cut-off scale where 
the effective theory ceases to be valid, and
\begin{equation}
{\cal L}_d=\sum_i C_i^{(d)} {\cal O}_i^{(d)}
\end{equation}
contains only (Lorentz and) gauge-invariant local operators, ${\cal O}_i^{(d)}$, of mass 
dimension $d$. In the so-called SM effective field theory (SMEFT), 
these operators are built using exclusively the SM symmetries and 
fields, assuming the Higgs belongs to an $SU(2)_L$ doublet. 
The Wilson coefficients, $C_i^{(d)}$, encode the dependence on the details of the
NP model. They can be obtained by matching with a particular ultraviolet (UV) completion
of the SM~\cite{delAguila:2000rc,delAguila:2008pw,delAguila:2010mx,deBlas:2014mba,Drozd:2015rsp,delAguila:2016zcb,Henning:2016lyp,Ellis:2016enq,Fuentes-Martin:2016uol,Zhang:2016pja,Ellis:2017jns,Wells:2017vla,deBlas:2017xtg}, 
allowing to project the EFT results into constraints on definite scenarios. 

At any order in the effective Lagrangian expansion a complete basis of physically independent operators contains only a
finite number of higher-dimensional interactions. In particular, for NP in the multi-TeV region, the precision of 
current EW measurements only allows to be sensitive to the leading terms in the $1/\Lambda$ expansion in  Eq.~(\ref{eq:EffLag}), i.e., 
the {\it dimension-six effective Lagrangian} (at dimension five there is only the Weinberg operator giving Majorana masses
to the SM neutrinos, which plays a negligible role in EW processes).
The first complete basis of independent dimension-six operators was introduced by Grzadkowski, Iskrzynski, Misiak, and Rosiek
and contains a total of 59 independent operators, barring flavour indices and Hermitian conjugates \cite{Grzadkowski:2010es}. 
This is what is commonly known in the literature as the {\it Warsaw} basis.

The main implementation of the dimension-six Standard Model effective Lagrangian in  \HEPfit is based in the Warsaw basis, 
though other operators outside this basis are also available for some calculations. Currently, all the dimension-six interactions 
entering in the EWPO as well as Higgs signal strengths have been included in the
{\tt NPSMEFTd6} model class. Two options are available, depending on whether lepton and quark flavour universality is 
assumed ({\tt NPSMEFTd6\_LFU\_QFU}) or not ({\tt NPSMEFTd6}). These implementations assume that we use the 
$\left\{M_Z, \alpha, G_F\right\}$ scheme for the SM EW input parameters. 
%, with $G_\mu$ the Fermi constant as extracted from $\mu$ decay. 
The complete list of operators as well the corresponding names for the \HEPfit model parameters can be 
found in the online documentation along with a complete description of the model flags.\footnote{The free parameters in the 
model also include several nuisance parameters to control theory uncertainties in certain Higgs processes.}

By default, the theoretical predictions for the experimental observables including the NP contributions coming from the effective Lagrangian are computed consistently with the assumption of only dimension-six effects. In other words, for a given observable, $O$, only effects of order $1/\Lambda^2$ are considered, and all NP contributions are linear in the NP parameters:
\begin{equation}
O=O_{\mathrm{SM}}+ \sum_i F_i \frac{C_i}{\Lambda^2}.
\label{eq:Odim6}
\end{equation}
Note that these linear contributions always come from the interference with the SM amplitudes. 
While this default behaviour is, in general, the consistent way to compute corrections in the effective Lagrangian expansion, there is no restriction in the code that forbids going beyond this level of approximation. In fact, further releases of the code are planned to also include the quadratic effects from the dimension-six interactions. The flag {\tt QuadraticTerms} will allow to test such effects.

%%%%%%%%%%%%%%%%%%%%%%%%
\subsection{Modified Higgs couplings in the \texorpdfstring{\boldmath$\kappa$}{kappa}-framework}
\label{sec:HiggsKappa}
%%%%%%%%%%%%%%%%%%%%%%%%

 \begin{table}[tb]
 \centering
  \begin{tabular}{| c c | c c | c c | c c |}
\hline
\textbf{Name}&
\textbf{Parameter}&
\textbf{Name}&
\textbf{Parameter}&
\textbf{Name}&
\textbf{Parameter}&
\textbf{Name}&
\textbf{Parameter}\\
\hline
 {\tt Kw}&
$\kappa_W$&
{\tt Kz}&
$\kappa_Z$&
{\tt Kg}&
$\kappa_g$&
{\tt Kga}&
$\kappa_\gamma$\\
{\tt Kzga}&
$\kappa_{Z\gamma}$&
 {\tt Ku}&
$\kappa_u$&
 {\tt Kc}&
$\kappa_c$&
 {\tt Kt}&
$\kappa_t$\\
 {\tt Kd}&
$\kappa_d$&
 {\tt Ks}&
$\kappa_s$&
 {\tt Kb}&
$\kappa_b$&
 {\tt Ke}&
$\kappa_e$\\
 {\tt Kmu}&
$\kappa_\mu$&
 {\tt Ktau}&
$\kappa_\tau$&
 {\tt BrHinv}&
${\rm BR}_{\rm inv}$&
 {\tt BrHexo}&
${\rm BR}_{\rm exo}$\\
\hline
  \end{tabular}
  \caption{Model parameters in the {\tt HiggsKigen} class. ``Name'' refers to the name of the parameter in \HEPfit that can be used in the configuration files.}\vspace{0.2cm}
 \label{tab:HiggsKigenpars}
\end{table}

In many scenarios of NP one of the main predictions are deviations in the Higgs boson couplings with respect to the SM ones. Such a scenario can be described in general by considering the following effective Lagrangian for a light Higgs-like scalar field $h$ \cite{Contino:2010mh,Azatov:2012bz}:
\begin{equation}
\begin{split}
{\cal L}=&\frac 12 \partial_\mu h~\! \partial^\mu h - V(h) + \frac{v^2}{4} \mathrm{Tr}(D_\mu \Sigma^\dagger D^\mu\Sigma)\left(1+2\kappa_V\frac{h}{v}+\dots\right)\\
&-m_{u^i} \left(\begin{array}{c c}\overline{u_L^i}&\overline{d_L^i}\end{array}\right) \Sigma \left(\begin{array}{c} {u_R^i}\\ 0 \end{array}\right) \left(1+ \kappa_u \frac{h}{v}+\dots \right) +\mathrm{h.c.}\\
&-m_{d^i} \left(\begin{array}{c c}\overline{u_L^i}&\overline{d_L^i}\end{array}\right) \Sigma \left(\begin{array}{c}0\\ {d_R^i} \end{array}\right) \left(1+ \kappa_d \frac{h}{v}+\dots \right) +\mathrm{h.c.}\\
&-m_{\ell^i} \left(\begin{array}{c c}\overline{\nu_L^i}&\overline{\ell_L^i}\end{array}\right) \Sigma \left(\begin{array}{c} 0\\ {\ell_R^i}\end{array}\right) \left(1+ \kappa_\ell \frac{h}{v}+\dots \right) +\mathrm{h.c.}~\!.
\label{eq:Lagh}
\end{split}
\end{equation}
This Lagrangian assumes an approximate custodial symmetry and the absence of other light degrees of freedom below the given cut-off scale. In the previous Lagrangian the longitudinal components of the $W$ and $Z$ gauge bosons, $\chi^a(x)$, are described by the $2\times 2$ matrix $\Sigma(x)=\mathrm{exp}\left( i \sigma_a \chi^a(x)/v\right)$, with $\sigma_a$ the Pauli matrices, and $V(h)$ is the scalar potential of the Higgs field, whose details are not relevant for the discussion here. The SM is recovered for $\kappa_V=\kappa_u=\kappa_d=\kappa_\ell=1$. 
Deviations in such a class of scenarios (and beyond) are conveniently encoded in the so-called $\kappa$ framework~\cite{LHCHiggsCrossSectionWorkingGroup:2012nn}. In this parameterization, deviations
from the SM in the Higgs properties are described by coupling modifier, $\kappa_i$, defined from the different Higgs production cross sections and decay widths. Schematically,
\begin{equation}
(\sigma \cdot {\rm BR})(i \to H \to f )=\kappa_i^2 \sigma^{\rm SM}(i\to H) \frac{\kappa_f^2 \Gamma^{\rm SM}(H\to f)}{\Gamma_H},
\end{equation}
where the total Higgs width, allowing the possibility of non-SM invisible or exotic decays, parameterized by ${\rm BR}_{\rm inv}$ and ${\rm BR}_{\rm exo}$, can be written as
\begin{equation}
\Gamma_H=\Gamma_H^{\rm SM} \frac{\sum_i \kappa_i^2 {\rm BR}_{i}^{\rm SM}}{1-{\rm BR}_{\rm inv}-{\rm BR}_{\rm exo}}.
\label{eq:HwidthK}
\end{equation}
The model class {\tt HiggsKigen} contains a general implementation of the parameterization described in the $\kappa$ framework, 
offering also several flags to adjust some of the different types of assumptions that are commonly used in the literature. 
The most general
set of coupling modifiers allowed in this class is described in Table~\ref{tab:HiggsKigenpars}, including also the possibility 
for non-SM contributions to invisible or exotic (non-invisible) Higgs decays.\footnote{As in the {\tt NPSMEFTd6} class, there are several nuisance parameters in the model to control theory uncertainties in certain Higgs processes. We refer to the documentation for a extensive list of the model parameters.}
Note that, even though the coupling modifiers are defined for all SM fermions, the current implementation of the code neglects modifications of the Higgs couplings to strange, up and down quarks, and to the electron. Furthermore, the parameters associated to $\kappa_{g,\gamma,Z\gamma}$, which are typically used in an attempt to interpret data allowing non-SM particles in the SM loops, are only meaningful if the model flag {\tt KiLoop} is active.

Finally, in scenarios like the one in Eq. (\ref{eq:Lagh}), while both $\kappa_V$ and $\kappa_f$ can modify the different Higgs production cross sections and decay widths, the leading corrections to EWPO come only from $\kappa_V$. These are given by the following 1-loop contributions to the oblique $S$ and $T$ parameters:
\begin{align}
S &= \frac{1}{12\pi} (1 - \kappa_V^2)
  \ln\bigg(\frac{\Lambda^2}{m_H^2}\bigg)\,,
&
T &= - \frac{3}{16\pi c_W^2} (1 - \kappa_V^2)
  \ln\bigg(\frac{\Lambda^2}{m_H^2}\bigg)\,,
\label{eq:ST}
\end{align}
where $\Lambda$ is the cutoff of the effective Lagrangian in Eq.~(\ref{eq:Lagh}).
We set $\Lambda = 4\pi v/\sqrt{|1-\kappa_V^2|}$, as given by the scale of violation of perturbative unitarity in $WW$ scattering. 

The above contributions from $\kappa_V$ to EWPO are also implemented in the {\tt HiggsKigen} class,
where $\kappa_V$ is taken from the model parameter associated to the $W$ coupling, $\kappa_W$.
Note however that, for $\kappa_W\not = \kappa_Z$ power divergences appear in the contributions to oblique corrections, 
and the detailed information of the UV theory is necessary for calculating the contributions to EWPO.
Therefore, in {\tt HiggsKigen} the use and interpretation of EWPO is subject to the use of the flag
{\tt Custodial}, which enables $\kappa_W = \kappa_Z$. 

Other flags in the model allow to use a global scaling for all fermion couplings (flag {\tt UniversalKf}), a global scaling for all SM couplings  (flag {\tt UniversalK}), and to trade the exotic branching ratio parameter by a scaling of the total Higgs width, according to Eq.~(\ref{eq:HwidthK}) (flag {\tt UseKH}).

%%%%%%%%%%%%%%%%%%%%%%%%%%%%%%%%%%%%%
\section{Some important observables implemented in \HEPfit}
\label{sec:Observables}
%%%%%%%%%%%%%%%%%%%%%%%%%%%%%%%%%%%%%
A large selection of observables has been implemented in
\HEPfit. Broadly speaking, these observables can be classified into
those pertaining to electroweak physics, Higgs physics, and flavour
physics. Observables should not necessarily be identified with
experimentally accessible quantities, but can also be used to impose
theoretical constraints, such as unitarity bounds, that can constrain
the parameter space of theoretical models, particularly beyond the SM.  In what
follows we give a brief overview of the main observables that are
available in \HEPfit along with some details about their
implementation when necessary.

%%%%%%%%%%%%%%%%%%%%%%%%%%%%%%%%%%%%%
\subsection{Electroweak physics}
\label{sec:EWPhysics}
%%%%%%%%%%%%%%%%%%%%%%%%%%%%%%%%%%%%%

The main EWPO have been implemented in
\HEPfit, including $Z$-pole observables as well as properties of
the $W$ boson (e.g. $W$ mass and decay width). 
%\HEPfit, including $Z$-pole and LEP II observables. 
The SM predictions for these observables are implemented including the
state-of-the-art of radiative corrections, following the work in
references~\cite{Sirlin:1980nh,Marciano:1980pb,Djouadi:1987gn,
  Djouadi:1987di,Kniehl:1989yc,Halzen:1990je,Kniehl:1991gu,Kniehl:1992dx,Barbieri:1992nz,Barbieri:1992dq,
  Djouadi:1993ss,Fleischer:1993ub,Fleischer:1994cb,Avdeev:1994db,Chetyrkin:1995ix,Chetyrkin:1995js,
  Degrassi:1996mg,Degrassi:1996ps,Degrassi:1999jd,
  Freitas:2000gg,vanderBij:2000cg,Freitas:2002ja,Awramik:2002wn,Onishchenko:2002ve,Awramik:2002vu,
  Awramik:2002wv,Awramik:2003ee,Awramik:2003rn,Faisst:2003px,Dubovyk:2016aqv,Dubovyk:2018rlg}. In
the current version of \HEPfit, all these observables are computed as
a function of the following SM input parameters: the $Z$, Higgs and
top-quark masses, $M_Z$, $m_h$ and $m_t$, respectively; the strong
coupling constant at the $Z$-pole, $\alpha_s(M_Z^2)$, and the
5-flavour contribution to the running of the electromagnetic constant
at the $Z$-pole, $\Delta \alpha_{\rm had}^{(5)}(M_Z^2)$.

The predictions including modifications due to NP effects are
also implemented for different models/scenarios, e.g. oblique
parameters
\cite{Kennedy:1988sn,Kennedy:1988rt,Peskin:1990zt,Altarelli:1990zd,Peskin:1991sw,Altarelli:1991fk,Altarelli:1993sz},
modified $Z$ couplings
\cite{Bamert:1996px,Haber:1999zh,Choudhury:2001hs,
  Agashe:2006at,Djouadi:2006rk,Kumar:2010vx,delAguila:2010mx,
  DaRold:2010as,Alvarez:2010js,Dermisek:2011xu,Djouadi:2011aj,
  Dermisek:2012qx,Batell:2012ca,Guadagnoli:2013mru}, the SMEFT
\cite{Buchmuller:1985jz,Grzadkowski:2010es}, etc. 

%%%%%%%%%%%%%%%%%%%%%%%%%%%%%%%%%%%%%
\subsection{Higgs physics}
\label{sec:HiggsPhysics}
%%%%%%%%%%%%%%%%%%%%%%%%%%%%%%%%%%%%%

In the Higgs sector, most of the observables currently included in \HEPfit are the Higgs-boson production cross sections or branching ratios, always normalized to the corresponding SM prediction. Modifications with respect to the SM are implemented for several models, e.g. {\tt NPSMEFTd6} or {\tt HiggsKigen}.
This set of observables allows to construct the different signal
strengths for each production$\times$decay measured at the LHC experiments and to test different NP hypotheses. 

Apart from the observables needed for LHC studies, the corresponding
observables for the production at future lepton colliders are also
implemented in \HEPfit. These are available for different values of centre-of-mass energies and/or polarization fractions, covering most of the options present in current proposals for such future facilities. Observables for studies at $ep$ colliders or at 100 TeV $pp$ colliders are also available in some cases.

%%%%%%%%%%%%%%%%%%%%%%%%%%%%%%%%%%%%%
\subsection{Flavour physics}
\label{sec:FlavourPhysics}
%%%%%%%%%%%%%%%%%%%%%%%%%%%%%%%%%%%%%
The list of observables already implemented in \HEPfit includes several
leptonic and semileptonic weak decays of flavoured mesons,
meson-antimeson oscillations, and lepton flavour and universality
violations. All these observables have been also implemented in models beyond the
SM.

\begin{table}[t!]
\centering
\begin{tabular}{|c|c|c|c|c|}
\hline
Processes 		& SM	& THDM 		& MSSM 		& $H_{eff}$\\
\hline
$\Delta F=2$		&\checkmark		&\checkmark
                                                &$\circ$
                                                                &\checkmark
  \\
$B\to\tau\nu$		&\checkmark		&\checkmark	&$\circ$		&$\circ$\\
$B\to D^{(*)}\ell\nu_{\ell}$	&\checkmark			&\checkmark
                                                &
                                                                &\checkmark
  \\
$B_{q}\to\mu\mu$	&\checkmark		&$\circ$		&$\circ$		&$\circ$\\
rare K decays		&$\circ$			&			&			&$\circ$\\
$B\to X_s\gamma$	&\checkmark		&\checkmark	&$\circ$		&\checkmark\\
$B\to V\gamma$	&\checkmark		&			&
                                                                &\checkmark
  \\
$B\to P/V\ell^+\ell^-$&\checkmark		&
                                                &
                                                                &\checkmark \\
$B\to X_s\ell^+\ell^-$&$\circ$			&			&			&$\circ$\\
$\ell_i\to \ell_j\gamma$	&				&			&\checkmark	&\\
$\ell_i\to 3\ell_j$		&				&			&\checkmark	&\\
$(g-2)_\mu$		&				&			&\checkmark	&\\
\hline
\end{tabular}
\caption{Some processes that have been implemented (\checkmark) or are under development ($\circ$) in \HEPfit for flavour physics.}
\label{tab:flav}
\end{table}%

\HEPfit has a dedicated flavour program in which several $\Delta F=2$,
$\Delta F=1$~\cite{Ciuchini:2015qxb,
  Ciuchini:2016weo,Cacchio:2016qyh,Ciuchini:2017mik,Ciuchini:2017gva,Gori:2017tvg,Kou:2018nap,Ciuchini:2018anp,Descotes-Genon:2018foz,Silvestrini:2018dos}
observables have been implemented to state-of-the-art precision in the SM and beyond. \HEPfit also includes observables
that require lepton flavour violation. In Table~\ref{tab:flav} we list
some of the processes that have either been fully implemented
(\checkmark) or are currently under development ($\circ$). We also
list out the models in which they have been implemented. $H_{\rm eff}$
refers here both to NP in the weak effective Hamiltonian as well as to
NP in the SMEFT. \HEPfit is continuously under development and the
list of available observables keeps increasing. The complete list can
be found in the \href{https://hepfit.roma1.infn.it/doc/latest-release/index.html}{online
  documentation}.

%%%%%%%%%%%%%%%%%%%%%%%%%%%%%%%%%%%%%
\subsection{Model-specific observables}
\label{sec:BSMPhysics}
%%%%%%%%%%%%%%%%%%%%%%%%%%%%%%%%%%%%%

Explicit NP models usually enlarge the particle spectrum, leading to
model-specific observables connected to (limits on) properties of new
particles (masses, production cross sections, etc.). Furthermore,
theoretical constraints such as vacuum stability, perturbativity,
etc. might be applicable to NP models. Both kinds of observables have
been implemented for several models extending the SM Higgs sector.

For example, in the THDM with a softly broken $Z_2$ symmetry
we implemented the conditions that the Higgs potential is bounded from
below at LO \cite{Deshpande:1977rw} and that the unitarity of
two-to-two scalar scattering processes is perturbative at NLO
\cite{Ginzburg:2005dt,Grinstein:2015rtl,Cacchio:2016qyh}. These
requirements can also be imposed at higher scales; the renormalization
group running is performed at NLO \cite{Chowdhury:2015yja}.  Also the
possibility that the Higgs potential features a second minimum deeper
than the electroweak vacuum at LO \cite{Barroso:2013awa} can be checked in
\HEPfit.  Similar constraints can be imposed on the Georgi-Machacek
model. In this case, both boundedness from below \cite{Arhrib:2011uy} and
unitarity \cite{Aoki:2007ah} are available at LO.

\begin{figure}[t!]
  \centering
  \includegraphics[width=0.4\textwidth]{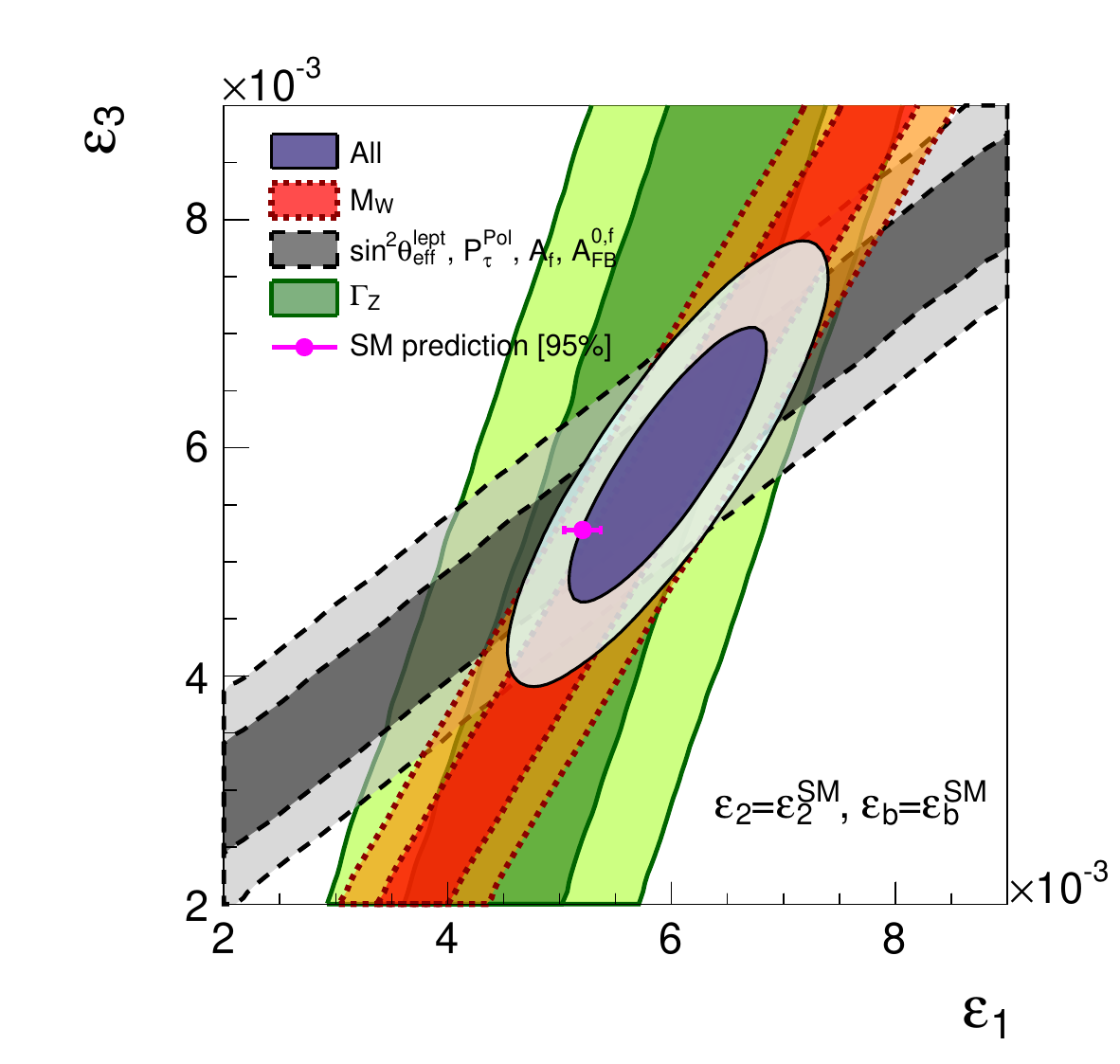}
  \includegraphics[width=0.4\textwidth]{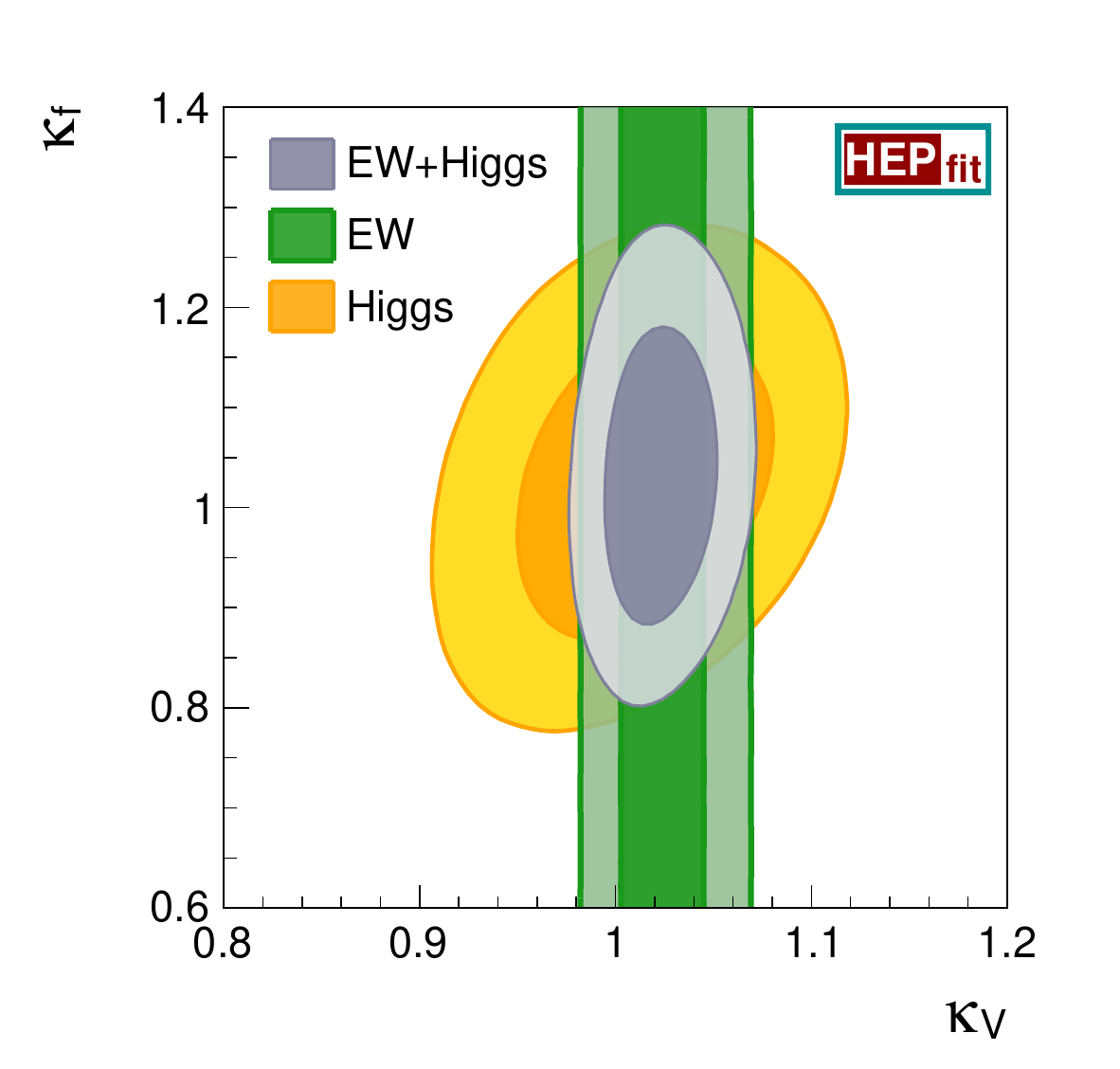}
  \includegraphics[width=0.5\textwidth]{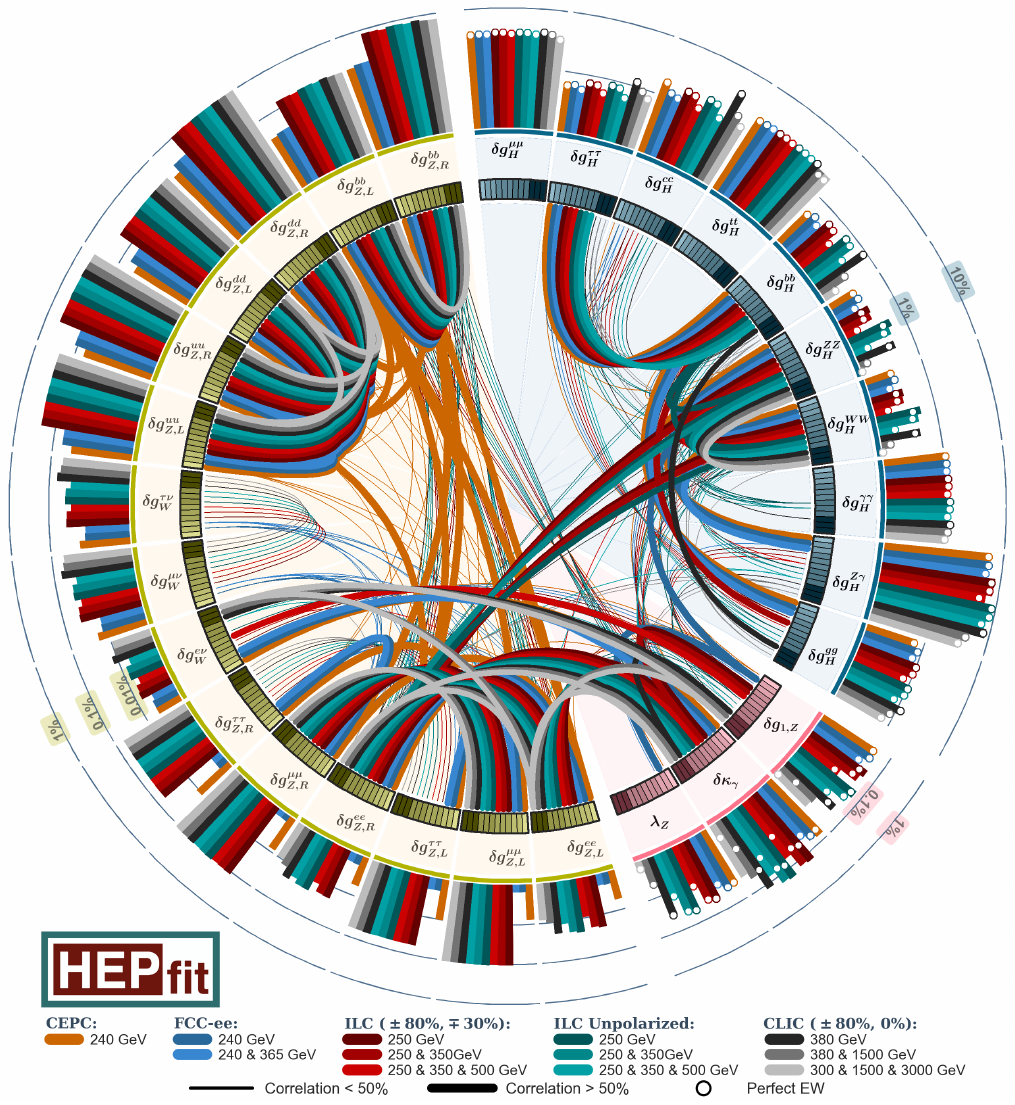}
  \caption{\it Selected results from the papers presented in Section
    \ref{sec:EWResults}. Top left (from ref.~\cite{Ciuchini:2013pca}):
    two-dimensional probability distribution for $\varepsilon_1$ and
    $\varepsilon_3$ in the fit, assuming $\varepsilon_2=\varepsilon_2^{\rm SM}$
    and $\varepsilon_b=\varepsilon_b^{\rm SM}$, showing the impact of
    different constraints. The SM prediction at 95\% is denoted by a
    point with an error bar. Top right (from
    ref.~\cite{deBlas:2016ojx}): two-dimensional $68\%$ (dark) and
    $95\%$ (light) probability contours for $\kappa_V$ and $\kappa_f$
    (from darker to lighter), obtained from the fit to the Higgs-boson
    signal strengths and the EWPO. Bottom (from
    ref.~\cite{deBlas:2019wgy}): a scheme-ball illustration of the
    correlations between Higgs and EW sector couplings.  The $Z$-pole
    runs are included for FCC-ee and CEPC.  Projections from
    HL-LHC and measurements from LEP and SLD are included in all
    scenarios.  The outer bars give the 1$\sigma$ precision on the
    individual coupling.}
  \label{fig:EWfigs}
\end{figure}

%%%%%%%%%%%%%%%%%%%%%%%%%%%%%%%%%%%%%
\section{Selected results using \HEPfit}
\label{sec:Results}
%%%%%%%%%%%%%%%%%%%%%%%%%%%%%%%%%%%%%
\HEPfit has so far been used to perform several analyses of
electroweak, Higgs and flavour physics in the SM and beyond. In this section we
highlight some of the results that have been obtained,
accompanied by a brief summary. The details of these analyses can be
found in the original publications.

%%%%%%%%%%%%%%%%%%%%%%%%%%%%%%%%%%%%%
\subsection{Electroweak and Higgs physics}
\label{sec:EWResults}
%%%%%%%%%%%%%%%%%%%%%%%%%%%%%%%%%%%%%

The first paper published using the nascent \HEPfit code featured a
full-fledged analysis of EWPO in the SM and beyond
\cite{Ciuchini:2013pca}, later generalized to include more NP models
and Higgs signal strengths
\cite{deBlas:2016ojx,deBlas:2018tjm,deBlas:2019wgy}. In the top left
plot of Figure \ref{fig:EWfigs}, taken from
ref.~\cite{Ciuchini:2013pca}, we show the two-dimensional probability
distribution for the NP parameters $\varepsilon_1$ and $\varepsilon_3$
\cite{Altarelli:1990zd,Altarelli:1991fk}, obtained assuming
$\varepsilon_2=\varepsilon_2^{\rm SM}$ and $\varepsilon_b=\varepsilon_b^{\rm
  SM}$. The impact of different constraints is also shown in the
plot. The top right plot, from ref.~\cite{deBlas:2016ojx}, presents the
results for modified Higgs couplings, multiplying SM Higgs couplings
to vector bosons by a universal scaling factor $\kappa_{V}$ and
similarly for fermions by a universal factor $\kappa_{f}$. The figure
shows the interplay between Higgs observables and EWPO in constraining
the modified couplings. The bottom plot, taken from
ref.~\cite{deBlas:2019wgy}, gives a pictorial representation of
constraints on several effective couplings, including correlations,
for different future lepton colliders.
The \HEPfit code was also used to obtain most of the results presented in the
future collider comparison study in Ref.~\cite{deBlas:2019rxi}. These are
summarized in the {\it Electroweak Physics} 
chapter of the {\it Physics Briefing Book} \cite{Strategy:2019vxc},
prepared as input for the Update of the
European Strategy for Particle Physics 2020.\footnote{\HEPfit was also
used for the combination studies in the individual analyses of the physics
potential of some of the different future 
collider projects, see Refs.~\cite{Azzi:2019yne,Cepeda:2019klc,Abada:2019zxq,Abada:2019lih}.}

%%%%%%%%%%%%%%%%%%%%%%%%%%%%%%%%%%%%%
\subsection{Flavour physics}
\label{sec:FlavourResults}
%%%%%%%%%%%%%%%%%%%%%%%%%%%%%%%%%%%%%

\begin{figure}[t]
\centering
\includegraphics[width=0.35\linewidth]{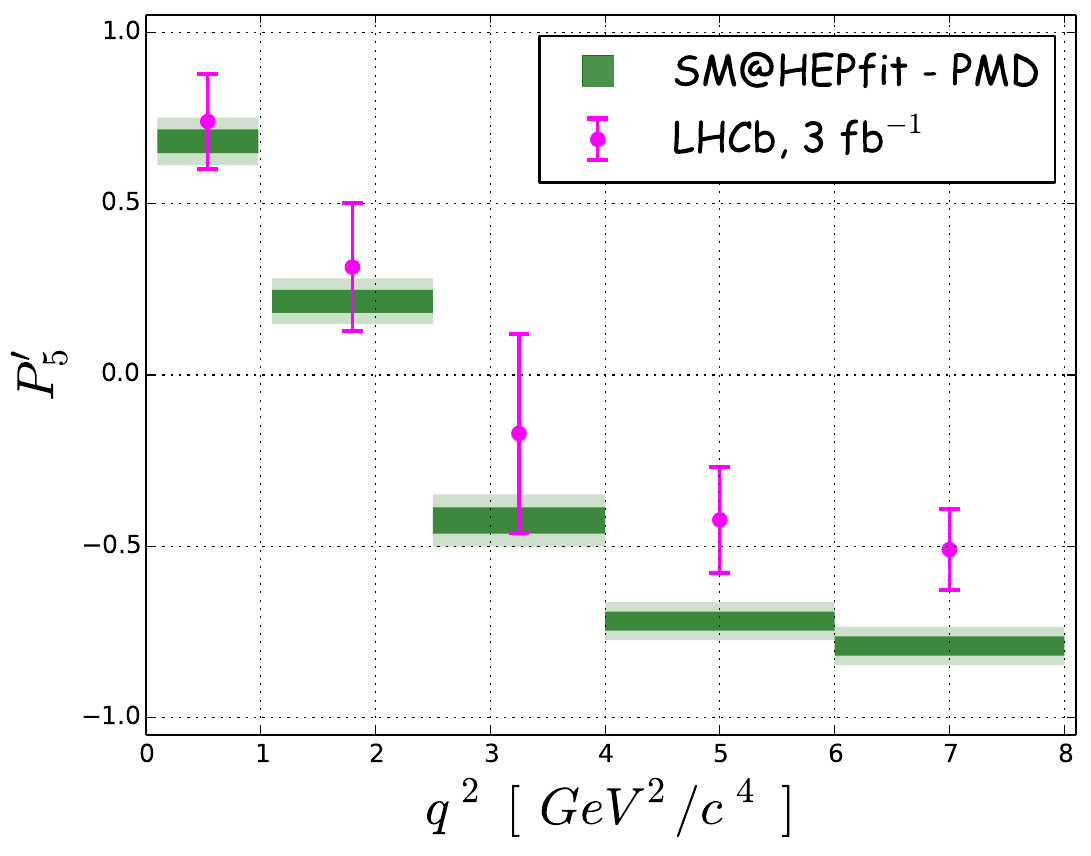}
\includegraphics[width=0.35\linewidth]{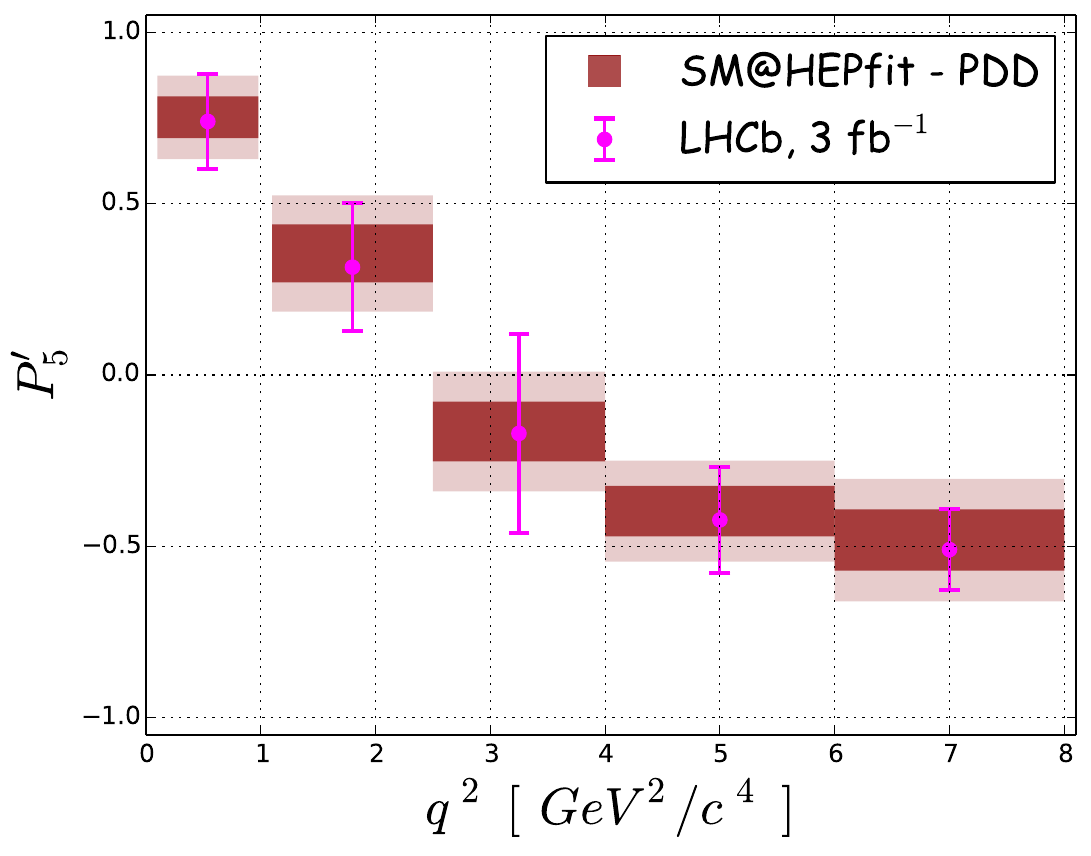}
\caption{\it Results of a fit for the LHCb results on the angular variable
  $P_{5}^{\prime}$ in two different theoretical scenarios: assuming
  the validity of an extrapolation of the QCD sum rules calculation of
  ref.~\cite{Khodjamirian:2010vf} at maximum hadronic recoil to the
  full kinematic range (left), or allowing for sizable long-distance
  contributions to be present for $q^{2}$ closer to $4 m_{c}^{2}$ (right). PMD
  refers to a more optimistic approach to hadronic contributions in $B\to K^*\ell^+\ell^-$
  decays and PDD refers to a more conservative apprach.
  For more details see ref.~\cite{Ciuchini:2018anp}}
\label{fig:plot_bsll}
\end{figure}

\begin{figure}[h!]
  \centering
  \includegraphics[width=0.8\textwidth]{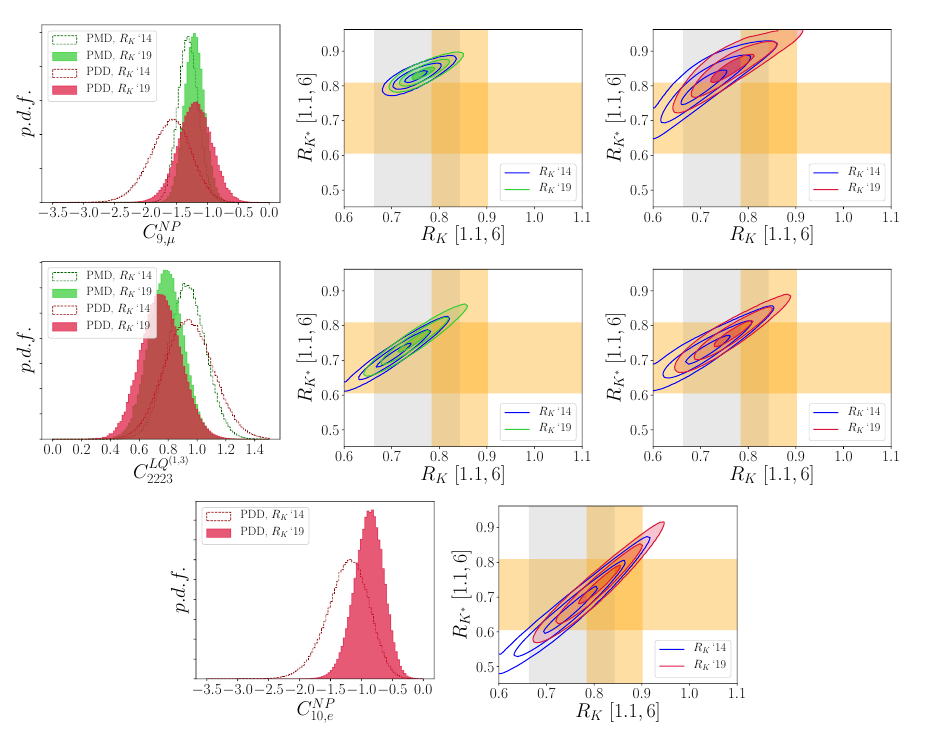}
  \vspace{-0.02\textwidth}
  \caption{\emph{First row: probability density function ({p.d.f.})
      for the NP contribution to the Wilson coefficient
      $C_{9,\mu}^{\rm NP}$. The green-filled {p.d.f.} shows the
      posterior obtained in the optimistic approach to hadronic
      contributions after the inclusion of the updated measurement for
      $R_K$, while the red-filled {p.d.f.} is the analogous posterior
      obtained allowing for sizable hadronic contributions (the dashed
      posteriors are the ones obtained employing the 2014 $R_K$
      measurement); the following panels report the combined 2D
      {p.d.f.} of the corresponding results for $R_K$ and $R_{K^*}$,
      where the colour scheme follows the one employed in the first
      panel. The horizontal band corresponds to the 1$\sigma$
      experimental region for $R_{K^*}$ from \cite{Aaij:2017vbb},
      while the two vertical bands corresponds to the previous and the
      current 1$\sigma$ experimental regions for $R_K$.  Second row:
      analogous to the first row, but relative to the SMEFT Wilson
      coefficient $C_{2223}^{LQ}$.  Third row: analogous to the first
      row, but relative to the NP contribution to the Wilson coefficient
      $C_{10,e}^{\rm NP}$. More details can be found
      in~\cite{Ciuchini:2019usw}}.}
  \label{fig:1D_mu_e}
\end{figure}

\begin{figure}
    \centering
    \includegraphics[width=\textwidth]{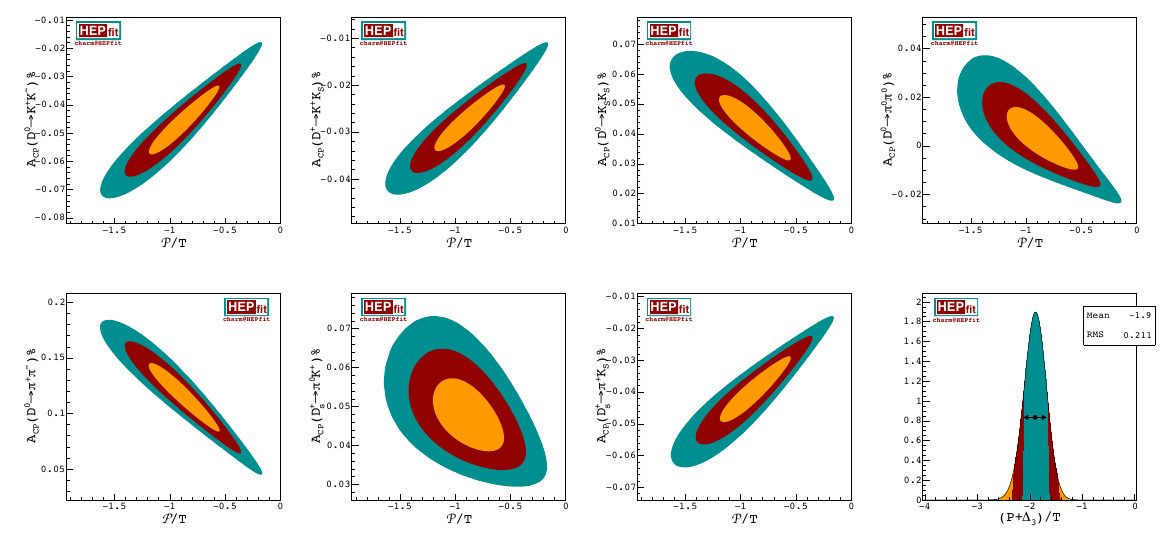}
    \caption{\it The correlations between the ratio of the penguin and tree contributions, $\mathpzcB{P}/T$, and the CP asymmetries (given in \%). HFLAV world average of $\Delta{A}_{\rm CP}$ has been used for the fit and these CP asymmetries correspond to the negative solution for the phases. The orange, red and green regions are the 68\%, 95\% and 99\% probability regions respectively. The bottom right-most panel shows the fit to  $(P+\Delta_3)/T=\mathpzcB{P}/T-1$. The orange, red and green regions are the 68\%, 95\% and 99\% probability regions respectively for the 2D histograms and the contrary for the 1D histogram. More details can be found in Ref.~\cite{Buccella:2019kpn}.}
    \label{fig:ch_fig}
\end{figure}

Analyses in flavour physics using \HEPfit has produced several results
following the claimed anomalies in $B$ physics. We started off by
reexamining the SM theoretical uncertainties and the possibility of
explaining the anomalies claimed in the angular distribution of
$B\to K^* \ell^+\ell^-$ decays through these
uncertainties~\cite{Ciuchini:2015qxb,Ciuchini:2016weo,Ciuchini:2017gva,Ciuchini:2018xll,Ciuchini:2018anp}. We
showed that the anomalies in the angular coefficients $P_5^\prime$
could be explained by allowing for a conservative estimate of the
theoretical uncertainties, see Figure~\ref{fig:plot_bsll}.

Having shown that the claimed deviations in the angular observables
from the SM predictions could be explained by making a more
conservative assumption about the non-perturbative contributions, we
addressed the cases for the deviations from unity of the measured
values of the lepton non-universal observables $R_{K^{(*)}}$, fitting
simultaneously for the NP Wilson coefficients and the
non-perturbative hadronic
contributions~\cite{Ciuchini:2017mik,Ciuchini:2019usw}. As before, we
studied the impact of hadronic contributions on the global fit. The
conclusions from our study were quite clear: on one hand the flavour
universal effects could be explained by enlarged hadronic effects,
reducing the significance of flavour universal NP effects. On the
other hand, flavour non-universal effects could only be explained by
the presence of NP contributions. In Figure~\ref{fig:1D_mu_e} we
present some of our results. More details can be found in
Ref.~\cite{Ciuchini:2019usw}.

Besides $B$ physics, \HEPfit has also been used for the analysis of
final state interactions (FSI) and CP asymmetries in $D\to PP$
($P=K,\pi$) decays. These have recently come to the forefront of
measurements with the pioneering 5$\sigma$ observation of
$\Delta A_{\rm CP} = A_{\mathrm{CP}} (D \to K^{+}K^{-}) -
A_{\mathrm{CP}} (D \to \pi^{+}\pi^{-})$ made by the LHCb
collaboration~\cite{Aaij:2019kcg,Betti:2019wlt,Pajero:2019jad}. This
work takes advantage of the high precision reached by the measurements
of the branching ratios in two particle final states consisting of
kaons and/or pions of the pseudoscalar charmed particles to deduce
the predictions of the SM for the CP violating asymmetries
in their decays. The amplitudes are constructed in agreement with the
measured branching ratios, where the ${SU}(3)_F$ violations
come mainly from the FSI and from the non-conservation of the
strangeness changing vector currents. A fit is performed of the
parameters to the branching fractions and $\Delta {A}_{\rm CP}$
using \HEPfit and predict several CP asymmetries using our
parameterization. In Figure~\ref{fig:ch_fig} the fit to the penguin
amplitude and the predictions for the CP asymmetries are shown. More
details can be found in Ref.~\cite{Buccella:2019kpn}.

%%%%%%%%%%%%%%%%%%%%%%%%%%%%%%%%%%%%%
\subsection{Constraints on specific new physics models: the case of extended scalar sectors}
\label{sec:BSMResults}
%%%%%%%%%%%%%%%%%%%%%%%%%%%%%%%%%%%%%
\begin{figure}[t]
    \centering
    \includegraphics[scale=0.47]{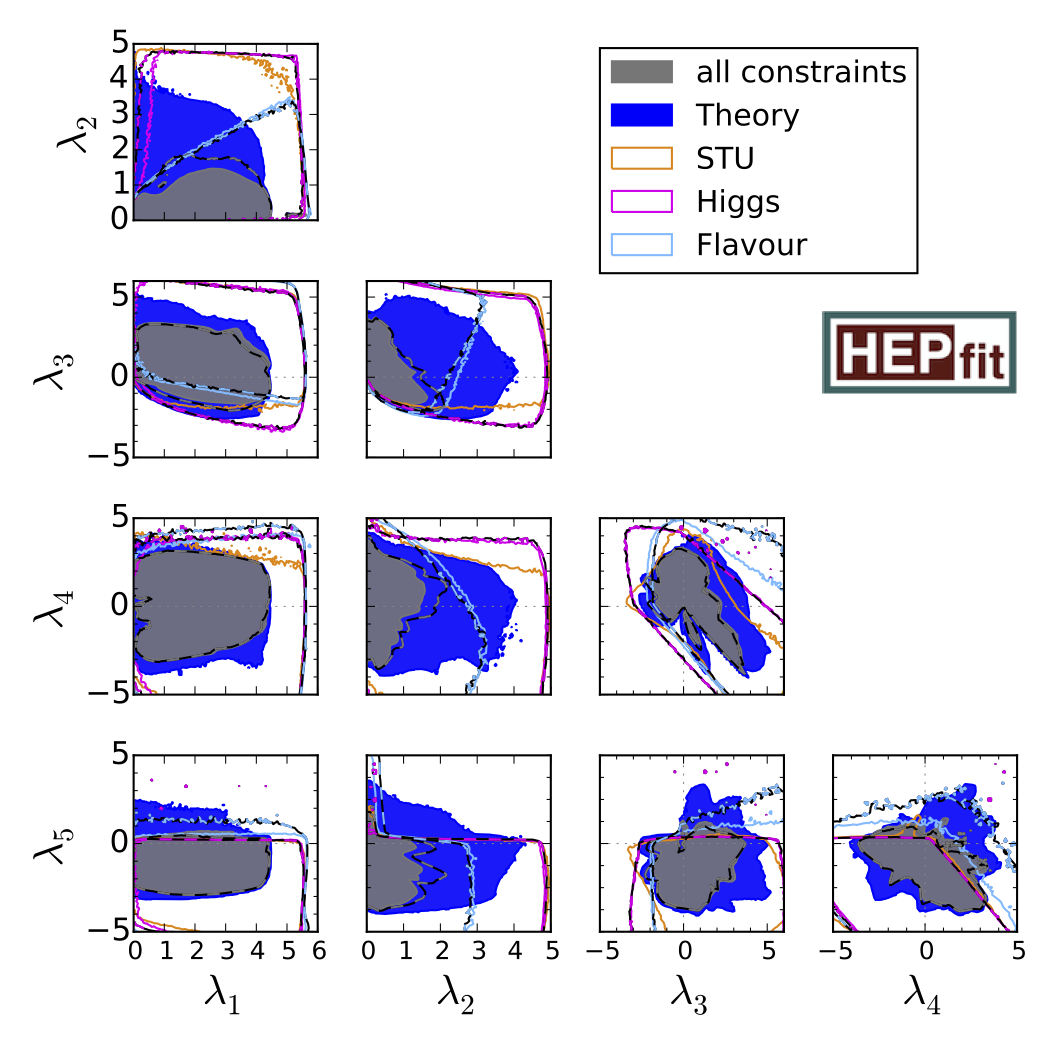}
    \includegraphics[scale=0.33]{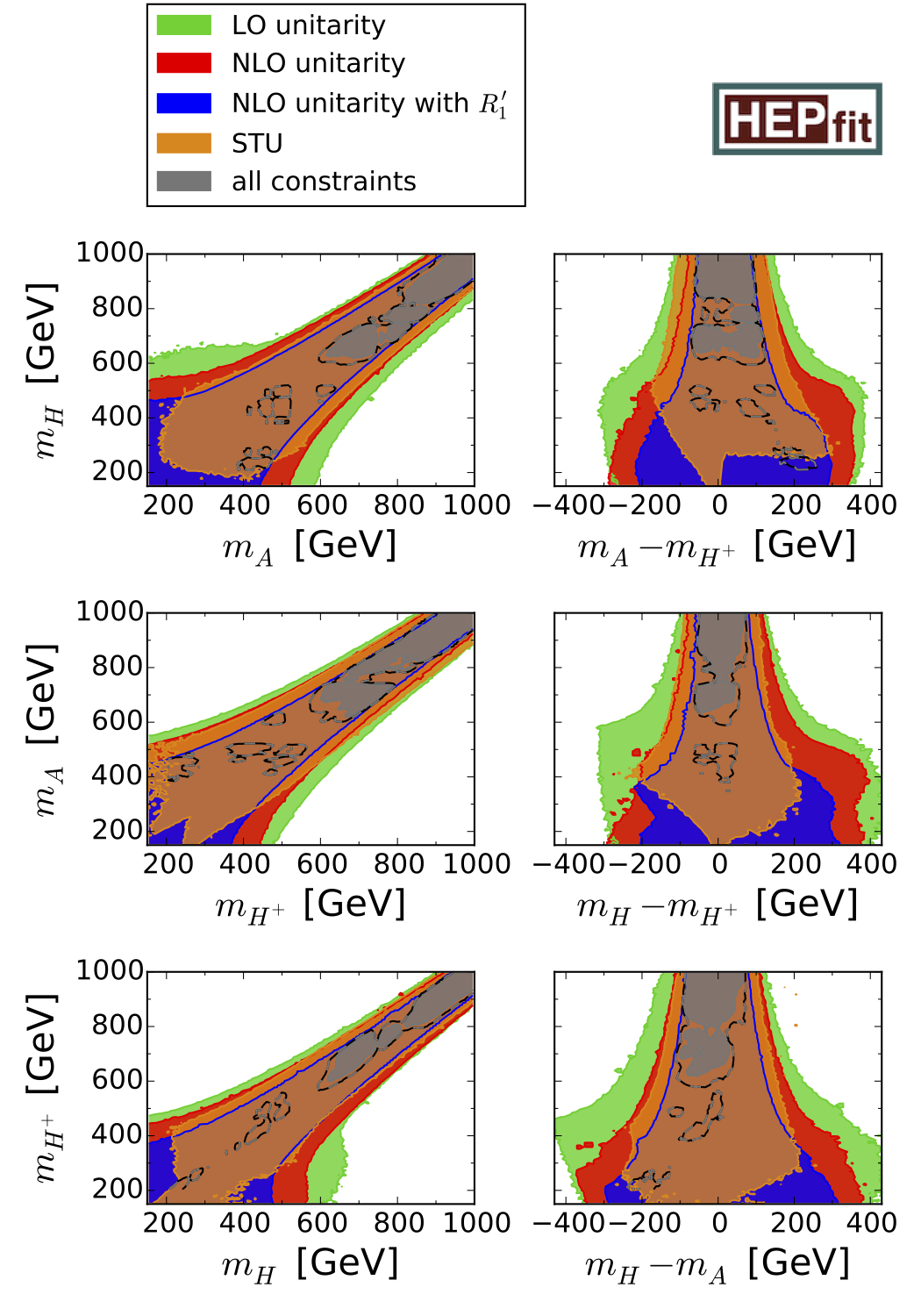}
    \caption{\it Left panel: $\lambda_i$ vs $\lambda_j$ planes. The blue
      shaded regions are 99.7\% probability areas taking into account
      theoretical constraints described in Section
      \ref{sec:BSMPhysics}. Orange, pink and light blue lines mark the
      95.4\% boundaries of fits using only the oblique parameters
      (STU), all Higgs observables (strengths and direct searches) and
      flavour observables, respectively. The grey contours are
      compatible with all theoretical and experimental bounds at a
      probability of 95.4\%. The solid lines are understood as the
      type II contours, the coloured dashed lines represent the
      corresponding type I fits. Right panel: Allowed regions in the
      heavy Higgs boson masses and their mass differences planes in
      the THDM of type I (dashed lines) and type II (solid lines). The
      unitarity bounds to the green, red and blue regions are meant at
      a probability of 99.7\%, and the orange and grey lines mark the
      95.4\% boundaries. More details can be found
      in~\cite{Cacchio:2016qyh}.  }
    \label{fig:2hdm_results_1}
\end{figure}
Several scalar extensions of the SM have been analysed using
\HEPfit. The THDM with a softly broken $Z_2$ symmetry has been
widely studied taking into account most of the relevant constraints available at the moment. Theoretical constraints described in Section
\ref{sec:BSMPhysics} are very useful to restrict the NP
parameter space. In this case approximate expressions for the NLO
perturbative unitarity conditions were obtained following the method
described in \cite{Murphy:2017ojk}. These expressions are valid in
the large center-of-mass limit and therefore they are only considered above a certain energy scale, default value of which is set to $750$ GeV. As shown in the left panel of Figure~\ref{fig:2hdm_results_1}, constraints on the $\lambda_i-\lambda_j$ (see Eq.~(\ref{eq:THDMVH})) planes can be obtained. These can be translated into constraints on physical observables such as the mass splitting of the scalar particles, $m_H-m_A, m_H-m_{H^{\pm}}$ and
$m_A - m_{H^{\pm}}$, as shown in the right panel of Figure~\ref{fig:2hdm_results_1}.  Theoretical constraints are independent of the specific model (type I, II, X, Y), which makes them especially useful.

Constraints on the mass planes coming from theoretical
observables are complementary to the oblique STU parameters described
in Section~\ref{sec:EWPhysics} (see left panel of
Figure~\ref{fig:2hdm_results_1}).  Results coming from the Higgs
observables (Section~\ref{sec:HiggsPhysics}) are of special interest
since they can provide us with direct bounds on the alignment angle
$\beta - \alpha$. Lastly, flavour observables described in
Section~\ref{sec:FlavourPhysics} provide bounds on the Yukawa
couplings (see Table~\ref{tab:THDMtypes}), which depend on the quantity $\tan \beta$ when written in the physical basis.

%Several scalar extensions of the SM were analysed using HEPfit: For the Two-Higgs-Doublet model (2HDM) with a softly broken $Z_2$ symmetry, most of the relevant constraints are available and were used for global fits \cite{Cacchio:2016qyh,Eberhardt:2017ulj,Chowdhury:2017aav,Eberhardt:2018lub}.
%Also a more general 2HDM with additional freedom in the Yukawa sector was analysed \cite{Gori:2017tvg}.
%Furthermore, the Georgi-Machacek model parameters were fitted to theoretical and LHC constraints \cite{Chiang:2018cgb} and the impact of different theoretical bounds on the 
%Manohar-Wise model with and without a second Higgs doublet was studied \cite{Cheng:2018mkc}.

%%%%%%%%%%%%%%%%%%%%%%%%%%%%%%%%%%%%%
\section{Installation}
\label{sec:Installation}
%%%%%%%%%%%%%%%%%%%%%%%%%%%%%%%%%%%%%

The installation of \HEPfit requires the availability of
\texttt{CMake} in the system. A description of \texttt{CMake} and the
details of how to install it can be found in the
\href{https://cmake.org/}{\texttt{CMake} website}. Most package
managers for Linux distributions should have a \texttt{CMake} package
available for installation. For Mac users, it can be either installed
from source or from a Unix port like
\href{https://www.macports.org/}{Darwin Ports} or
\href{http://www.finkproject.org/}{Fink}, or the installation package
can be downloaded from the \href{https://cmake.org/}{\texttt{CMake}
  website}. We list below the dependencies that need to be satisfied to
successfully install \HEPfit:

\begin{itemize}
\item {\bf \texttt{GSL}:}  The GNU Scientific Library (\texttt{GSL})
  is a \texttt{C} library for numerical computations. It can be found
  on the \href{http://www.gnu.org/software/gsl/}{GSL website}. Most
  Linux package managers will have a stable version as will any ports
  for Mac. \HEPfit is compatible with \texttt{GSL} v1.16 or greater. 

\item {\bf \ROOT v5 or greater:}  \ROOT is an object
  oriented data analysis framework. You can obtain it from the
  \href{http://root.cern.ch/}{\ROOT website}. \BAT requires
  \ROOT v5.34.19 or greater. Both \HEPfit and \BAT are
  compatible with \ROOT v6.  {\bf NOTE:} If \texttt{GSL} is
  installed before compiling \ROOT from source, then
  \ROOT builds by default the MathMore library, which depends
  on \texttt{GSL}. Hence it is recommended to install \texttt{GSL}
  before installing \ROOT.

\item {\bf \texttt{BOOST}:} \texttt{BOOST} is a \texttt{C++} library
  which can be obtained from the
  \href{http://www.boost.org}{\texttt{BOOST} website} or from Linux
  package managers or Mac ports. \HEPfit only requires the
  \texttt{BOOST} headers, not the full libraries, so a header-only
  installation is sufficient. \HEPfit has been tested to work with
  \texttt{BOOST} v1.53 and greater.

\item {\bf MPI:} Optionally, \HEPfit can be compiled with MPI for
  usage in parallelized clusters and processors supporting
  multi-threading. In this case, the \HEPfit installer will patch and
  compile \BAT with MPI support as described below. To this purpose
  one needs \href{https://www.open-mpi.org/}{\texttt{OpenMPI}} which is also
  available through package managers in Linux and ports on Mac.

\item {\bf \BAT v1.0 (not required for the Library mode):} The 
    \href{https://www.mppmu.mpg.de/bat/}{\BAT website} offers the source code for \BAT but it should {\em not} be used with \HEPfit since a patch is required to integrate \BAT with \HEPfit. With the compilation 
    option \texttt{-DBAT\_INSTALL=ON} explained below, the \HEPfit installation package 
    will download, patch and install \BAT. The parallelized version of \BAT compatible with the parallelized version of \HEPfit\ can be installed with the additional option \texttt{-DMPIBAT=ON} for which MPI must be installed (see ``MPI Support'' below).

\end{itemize}
%%%%%%%%%%%%%%%%%%%%%%%%%%%%%%%%%%%%
\subsection{Installation procedure}
%%%%%%%%%%%%%%%%%%%%%%%%%%%%%%%%%%%%

\noindent {\bf Quick Installation Instructions:}

In a nutshell, if all dependencies are satisfied, for a fully MPI
compatible MCMC capable \HEPfit version \texttt{x.y} installation from
the tarball downloaded from the \href{https://hepfit.roma1.infn.it/}{
  \HEPfit website}:

\begin{lstlisting}[numbers=none,language=]
$ tar xvzf HEPfit-x.y.tar.gz
$ mkdir HEPfit-x.y/build 
$ cd HEPfit-x.y/build 
$ cmake .. -DLOCAL_INSTALL_ALL=ON -DMPIBAT=ON
$ make
$ make install
\end{lstlisting}
To run your first example:
\begin{lstlisting}[numbers=none,language=]
$ cd examples/MonteCarloMode/
$ make  
$ mpiexec -n 5 ./analysis ../config/StandardModel.conf MonteCarlo.conf
\end{lstlisting} 
This is all you need for running a MCMC simulation on 5 cores with the model, parameters and observables specified in the configuration files in \texttt{examples/config} directory with \HEPfit. For variations please read what follows. \\\\
\noindent {\bf Detailed Installation Instructions}

Unpack the tarball containing the \HEPfit version \texttt{x.y} source
which you can obtain from the
\href{https://hepfit.roma1.infn.it/}{\HEPfit website}. A directory
called \texttt{HEPfit-x.y} will be created containing the source
code. To generate Makefiles, enter the source directory and run \texttt{CMake}:

\begin{lstlisting}[numbers=none,language=]
$ cd HEPfit-x.y  
$ cmake . <options>  
\end{lstlisting}

\noindent{\bf (RECOMMENDED:)} Alternatively, a directory separate from the source directory can be made for
building \HEPfit (recommended as it allows for easy deletion of the build):
\begin{lstlisting}[numbers=none,language=]
$ mkdir HEPfit-x.y/build  
$ cd HEPfit-x.y/build  
$ cmake .. <options>  
\end{lstlisting}
where the available options are:

\begin{itemize}
\item \texttt{-DLOCAL\_INSTALL\_ALL=ON}: to install \BAT and \HEPfit in the current directory (default: OFF). 
    This is equivalent to setting the combination of the options:
    
\begin{lstlisting}[numbers=none,language=]
-DCMAKE_INSTALL_PREFIX=./HEPfit -DBAT_INSTALL_DIR=./BAT -DBAT_INSTALL=ON
\end{lstlisting} 
These variables cannot be modified individually when \texttt{-DLOCAL\_INSTALL\_ALL=ON} is set. 

\item \texttt{-DCMAKE\_INSTALL\_PREFIX=<HEPfit installation directory>}: the directory in which \HEPfit will be installed (default: \texttt{/usr/local}).  
  
\item \texttt{-DNOMCMC=ON}: to enable the mode without MCMC (default: OFF).

\item \texttt{-DDEBUG\_MODE=ON}: to enable the debug mode (default: OFF).

\item \texttt{-DBAT\_INSTALL\_DIR=<BAT installation directory>}: (default: \texttt{/usr/local}). This option is overridden by \texttt{-DLOCAL\_INSTALL\_ALL=ON} .

\item \texttt{-DBAT\_INSTALL=ON} to download and install \BAT. This is relevant only if
\texttt{-DNOMCMC=ON} is not set. Use \texttt{-DBAT\_INSTALL=OFF} only if you know your \BAT installation
is already patched by \HEPfit and is with or without MPI support as needed. (default: ON).

\item \texttt{-DMPIBAT=ON}: to enable support for MPI for both \BAT and \HEPfit
    (requires an implementation of MPI, default: OFF).

\item \texttt{-DMPI\_CXX\_COMPILER=<path to mpi>/mpicxx}: You can specify the MPI compiler with this option.

\item \texttt{-DBOOST\_INCLUDE\_DIR=<boost custom include path>/boost/}: if BOOST is not installed in the search path then you can specify where it is with this option. The path must end with the \texttt{boost/} directory which contains the headers.

\item \texttt{-DGSL\_CONFIG\_DIR=<path to gsl-config>}: \HEPfit used \texttt{gsl-config} to get the GSL parameters. If this is not in the search path, you can specify it with this option. 

\item \texttt{-DROOT\_CONFIG\_DIR=<path to root-config>}: \HEPfit used \texttt{root-config} to get the \ROOT parameters. If this is not in the search path, you can specify it with this option. 

\item \texttt{-DINTEL\_FORTRAN=ON}: If you are compiling with INTEL compilers then this flag turns
on support for the compilers (default: OFF). 

\end{itemize}
Setting the option \texttt{-DBAT\_INSTALL=ON}, the \HEPfit installer will download, 
compile and install the \BAT libraries.\\\\
{\bf NOTE:}
If \BAT libraries and headers are present in target directory for \BAT they will be overwritten unless \texttt{-DBAT\_INSTALL=OFF} is set. This is done so that the correct patched version of \BAT compatible with
\HEPfit gets installed.
{\bf No MCMC mode:}
The generated Makefiles are used for building a \HEPfit library. If
you do not perform a Bayesian statistical analysis with the MCMC, you can use the option \texttt{-DNOMCMC=ON}. In
this case, BAT is not required. \\\\
{\bf MPI Support:}
If you want to perform an MCMC run with MPI support, you can specify
the option \texttt{-DMPIBAT=ON}. This option must not be accompanied with
\texttt{-DBAT\_INSTALL=OFF} in order to enable the \HEPfit installer to
download, patch and compile BAT and build \HEPfit with MPI support:
\begin{lstlisting}[numbers=none,language=]
$ cmake . -DMPIBAT=ON <other options>  
\end{lstlisting}
\vspace*{3mm}
{\bf ROOT:}
\texttt{CMake} checks for \ROOT availability in the system and fails if \ROOT is
not installed. You can specify the path to \texttt{root-config} using the
option \texttt{-DROOT\_CONFIG\_DIR=<path to root-config>}. \\\\
{\bf BOOST:}
\texttt{CMake} also checks for \texttt{BOOST} headers availability in the system and fails if
\texttt{BOOST} headers are not installed. You can specify the path to the \texttt{BOOST} include
files with \texttt{-DBOOST\_INCLUDE\_DIR=<boost custom include path>/boost/}. \\\\
The {\bf recommended installation} flags for a locally installed \HEPfit with full MPI and MCMC support is:
\begin{lstlisting}[numbers=none,language=]
$ cmake . -DLOCAL_INSTALL_ALL=ON -DMPIBAT=ON
\end{lstlisting}
This will enable easy portability of all codes and easy upgrading to future version as nothing will be installed system wide.  Also, this is useful if you do not have root access and cannot install software in system folders. After successful \texttt{CMake} run, execute the build commands:

\begin{lstlisting}[numbers=none,language=]
$ make  
$ make install  
\end{lstlisting} 
to compile and install \HEPfit, where the command \texttt{make VERBOSE=1}
enables verbose output and \texttt{make -j} allows for parallel compilation.
Note that depending on the setting of installation prefix you might
need root privileges to be able to install \HEPfit with \texttt{sudo make
install} instead of just \texttt{make install}.\\

%%%%%%%%%%%%%%%%%%%%%%%%%%%%%%%%%
\subsection{Post installation}
\label{sec:postInstall}
%%%%%%%%%%%%%%%%%%%%%%%%%%%%%%%%%

After the completion of the installation with \texttt{make install} the following three files can be found in the installation location. The file \texttt{libHEPfit.h} is a combined header file corresponding to the library \texttt{libHEPfit.a}.

\begin{itemize}
\item[]{\bf Executable:} \texttt{<CMAKE\_INSTALL\_PREFIX>/bin/hepfit-config}
\item[]{\bf Library:} \texttt{<CMAKE\_INSTALL\_PREFIX>/lib/libHEPfit.a}
\item[]{\bf Combined Header:} \texttt{<CMAKE\_INSTALL\_PREFIX>/include/HEPfit/HEPfit.h}
\end{itemize}

\noindent{\bf Using hepfit-config:}
A hepfit-config script can be found in the \texttt{<CMAKE\_INSTALL\_PREFIX>/bin/}
directory, which can be invoked with the following options:
\begin{itemize}
\item \texttt{--cflags} to obtain the include path needed for compilation against the \HEPfit library.

\item \texttt{--libs} to obtain the flags needed for linking against the \HEPfit library.
\end{itemize}

\noindent{\bf Examples:}
The example programs can be found in the \HEPfit build directory:  
\begin{itemize}
\item \texttt{examples/LibMode\_config/}  
\item \texttt{examples/LibMode\_header/} 
\item \texttt{examples/MonteCarloMode/}
\item \texttt{examples/EventGeneration/}
\item \texttt{examples/myModel/}
\end{itemize}
The first two demonstrate the usage of the \HEPfit library, while 
the third one can be used for testing a Monte Carlo run with the \HEPfit 
executable. The fourth example can be used to generate values of observables with a sample of parameters drawn from the parameter space. The fifth one is an example implementation of a custom 
model and custom observables. To make an executable to run these examples:
\begin{lstlisting}[numbers=none,language=]
$ cd examples/MonteCarloMode/
$ make  
\end{lstlisting} 
This will produce an executable called \texttt{analysis} in the current directory that can be used to run \HEPfit. The details are elaborated on in the next section.

%%%%%%%%%%%%%%%%%%%%%%%%%%%%%%%%%%%%%
\section{Usage and examples}
\label{sec:Usage}
%%%%%%%%%%%%%%%%%%%%%%%%%%%%%%%%%%%%%
After the \HEPfit installer generates the 
library \texttt{libHEPFit.a} along with header files included in a combined
header file, \texttt{HEPfit.h}, the given example implementation can be used 
to perform a MCMC based Bayesian statistical analysis.
Alternatively, the library can be used to obtain predictions
of observables for a given point in the parameter space of a model, 
allowing \HEPfit to be called from the user's own
program. We explain both methods below. In addition \HEPfit provides
the ability to the user to define custom models and observables as
explained in \ref{sec:mymodel}. We give a brief description on how to
get started with custom models and observables.

%%%%%%%%%%%%%%%%%%%%%%%%%%%%%%%%%%%%%
\subsection{Monte Carlo mode}
\label{sec:MC}
%%%%%%%%%%%%%%%%%%%%%%%%%%%%%%%%%%%%%
The Monte Carlo analysis is performed with the \BAT library. First,
a text configuration file (or a set of files) containing a list of model parameters,
model flags and observables to be analyzed has to be prepared. Another configuration
file for the Monte Carlo run has to be prepared, too.\\\\
{\bf \large Step 1: Model configuration file}\\

The configuration files are the primary way to control the behaviour of
the code and to detail its input and output. While a lot of checks
have been implemented in \HEPfit to make sure the configuration files
are of the right format, it is not possible to make it
error-proof. Hence,  care should be taken in preparing these files. A configuration file for model parameters, model flags, and
observables is written as follows:\\

\begin{lstlisting}[language=]
StandardModel
# Model parameters:
ModelParameter  mtop        173.2       0.9         0.  
ModelParameter  mHl         125.6       0.3         0.  
...
CorrelatedGaussianParameters   V1_lattice 2
ModelParameter  a_0V    0.496   0.067   0.
ModelParameter  a_1V   -2.03    0.92    0.
1.00    0.86
0.86    1.00

<All the model parameters have to be listed here>

# Observables:
Observable  Mw        Mw        M_{W}      80.3290 80.4064 MCMC weight 80.385 0.015 0.  
Observable  GammaW    GammaW    #Gamma_{W} 2.08569 2.09249 MCMC weight 2.085  0.042 0.  
#
# Correlated observables:
CorrelatedGaussianObservables Zpole2 7  
Observable  Alepton   Alepton   A_{l}      0.143568 0.151850 MCMC weight 0.1513  0.0021  0.  
Observable  Rbottom   Rbottom   R_{b}      0.215602 0.215958 MCMC weight 0.21629 0.00066 0.  
Observable  Rcharm    Rcharm    R_{c}      0.172143 0.172334 MCMC weight 0.1721  0.0030  0.  
Observable  AFBbottom AFBbottom A_{FB}^{b} 0.100604 0.106484 MCMC weight 0.0992  0.0016  0.  
Observable  AFBcharm  AFBcharm  A_{FB}^{c} 0.071750 0.076305 MCMC weight 0.0707  0.0035  0.  
Observable  Abottom   Abottom   A_{b}      0.934320 0.935007 MCMC weight 0.923   0.020   0.  
Observable  Acharm    Acharm    A_{c}      0.666374 0.670015 MCMC weight 0.670   0.027   0.  
1.00   0.00   0.00   0.00   0.00   0.09   0.05
0.00   1.00  -0.18  -0.10   0.07  -0.08   0.04
0.00  -0.18   1.00   0.04  -0.06   0.04  -0.06
0.00  -0.10   0.04   1.00   0.15   0.06   0.01
0.00   0.07  -0.06   0.15   1.00  -0.02   0.04
0.09  -0.08   0.04   0.06  -0.02   1.00   0.11
0.05   0.04  -0.06   0.01   0.04   0.11   1.00  
#
# Output correlations: 
Observable2D  MwvsGammaW Mw M_{W} 80.3290 80.4064 noMCMC noweight GammaW #Gamma_{W} 2.08569 2.09249  
...
Observable2D Bd_Bsbar_mumu noMCMC noweight
Observable   BR_Bdmumu      BR(B_{d}#rightarrow#mu#mu)    1. -1.  1.05e-10   0.   0.
Observable   BRbar_Bsmumu   BR(B_{s}#rightarrow#mu#mu)    1. -1.  3.65e-9    0.   0.
...
Observable2D S5_P5 noMCMC noweight
BinnedObservable   S_5      S_5    1. -1.  0.   0.   0.   4.   6.
BinnedObservable   P_5      P_5    1. -1.  0.   0.   0.   4.   6.
#
# Including other configuration files
IncludeFile Flavour.conf

\end{lstlisting}
where the lines beginning with the `\#' are commented out. Each line has to be written as follows: 

\begin{enumerate}
\item The first line must be {\bf the name of the model} to be analyzed,
   where the available models are listed in the \HEPfit online documentation.
   
\item {\bf Model flags}, if necessary, should be specified right after the model because some of them can control the way the input parameters are read.   
\begin{lstlisting}[numbers=none,language=]
ModelFlag <name> <value>
\end{lstlisting}
  
\item  {\bf A model parameter} is given in the format:
\begin{lstlisting}[numbers=none,language=]
ModelParameter <name> <central value> <Gaussian error> <flat error>
\end{lstlisting}
where all the parameters in a given model (see the online documentation) have
  to be listed in the configuration file.

\item {\bf A set of correlated model parameters} is specified with 
\begin{lstlisting}[numbers=none,language=]
CorrelatedGaussianParameters name Npar
\end{lstlisting}
   which initializes a set of \texttt{Npar} correlated parameters. It must be
   followed by exactly \texttt{Npar} ModelParameter lines and then by \texttt{Npar} rows of
   \texttt{Npar} numbers for the correlation matrix. See the example above.

\item  {\bf An \texttt{Observable}} to be computed is specified in one of the following formats:
\begin{lstlisting}[language=]
Observable <name> <obs label> <histolabel> <min> <max> (no)MCMC (no)weight    <central value> <Gaussian error> <flat error>
#
Observable <name> <obs label> <histolabel> <min> <max> (no)MCMC file          <filename> <histoname>
#
Observable <name> <obs label> <histolabel> <min> <max> noMCMC noweight
#
Observable <name> <obs label> <histolabel> <min> <max> noMCMC writeChain
\end{lstlisting}
\begin{itemize}
  \item {\bf \texttt{<name>}} is a user given name for different observables which must be unique for each observable.
  \item {\bf \texttt{<obs label>}} is the theory label of the
    observable (see the online documentation).
  \item {\bf \texttt{<histolabel>}} is used for the label of the output \ROOT histogram,
  while {\bf \texttt{<min>}} and {\bf \texttt{<max>}} represent the
  range of the histogram (if {\bf \texttt{<min>}} $\ge$ {\bf
    \texttt{<max>}} the range of the histogram is set automatically).
  \item  {\bf \texttt{(no)MCMC}} is the flag specifying whether the
    observable should be included in likelihood used for the MCMC sampling.
  \item {\bf \texttt{(no)weight}} specifies if the observable weight will be computed or not. If \texttt{weight} is specified with \texttt{noMCMC}
    then a chain containing the weights for the observable will be stored in the
    \texttt{MCout*.root} file.  
  \item {\bf \texttt{noMCMC noweight}} is the combination to be used
    to get a prediction for an observable.
  \item When the \texttt{weight} option is specified, at least one of
    the 
    \texttt{ <Gaussian error>} or the \texttt{<flat error>} must be
    non-vanishing, and the \texttt{<central value>} must of course be specified.
  \item When using the \texttt{file} option, a histogram in a
    \ROOT file must be specified by the name of the \ROOT file ({\bf \texttt{filename}}) and then the
    name of the histogram ({\bf \texttt{histoname}}) in the file
    (including, if needed, the directory).
  \item The \texttt{writeChain} option allows one to write all the values of the observable generated during the main run of the MCMC into the \ROOT file.
\end{itemize}

\item {\bf A \texttt{BinnedObservable}} is similar in construction to an \texttt{Observable} but with two extra arguments specifying the upper and lower limit of the bin:
\begin{lstlisting}[language=]
BinnedObservable <name> <obs label> <histolabel> <min> <max> (no)MCMC        (no)weight <central value> <Gaussian error> <flat error> <bin_min>       <bin_max>
#
BinnedObservable <name> <obs label> <histolabel> <min> <max> (no)MCMC        (no)weight <filename> <histoname> <bin_min> <bin_max>
#
BinnedObservable <name> <obs label> <histolabel> <min> <max> noMCMC writeChain 0. 0. 0. <bin_min> <bin_max>
\end{lstlisting}
Because of the order of parsing the \texttt{<central value> <Gaussian error> <flat error>} cannot be dropped out even in the \texttt{noMCMC noweight} case for a \texttt{BinnedObservable}.

\item {\bf A \texttt{FunctionObservable}} is the same as a \texttt{BinnedObservable} but with only one extra argument that points to the value at which the function is computed:
\begin{lstlisting}[language=]
FunctionObservable <name> <obs label> <histolabel> <min> <max> (no)MCMC        (no)weight <central value> <Gaussian error> <flat error> <x_value>
#
FunctionObservable <name> <obs label> <histolabel> <min> <max> (no)MCMC        (no)weight <filename> <histoname> <x_value>
#
FunctionObservable <name> <obs label> <histolabel> <min> <max> noMCMC writeChain 0. 0. 0. <x_value>
\end{lstlisting}

\item {\bf An asymmetric Gaussian} constraint can be set using \texttt{AsyGausObservable}:
\begin{lstlisting}[language=]
AsyGausObservable <name> <obs label> <histolabel> <min> <max> (no)MCMC          (no)weight <central value> <left_error> <right_error>
\end{lstlisting}

\item  {\bf Correlations among observables} can be taken into
   account with the line\\ \texttt{CorrelatedGaussianObservables name Nobs},
   which initializes a set of \texttt{Nobs} correlated observables. It must be
   followed by exactly \texttt{Nobs} \texttt{Observable} lines and then by \texttt{Nobs} rows of
   \texttt{Nobs} numbers for the correlation matrix (see the above example).
   One can use the keywords \texttt{noMCMC} and \texttt{noweight},
   instead of \texttt{MCMC} and \texttt{weight}.
\begin{lstlisting}[language=]
CorrelatedGaussianObservables <name> Nobs
Observable <name> <obs label> <histolabel> <min> <max> (no)MCMC (no)weight    <central value> <Gaussian error> <flat error>
Observable <name> <obs label> <histolabel> <min> <max> (no)MCMC (no)weight    <central value> <Gaussian error> <flat error>
...
...
<Total of Nobs lines of Observables>
<Nobs%*$\times$Nobs correlation matrix>
\end{lstlisting}
Any construction for \texttt{Observable} mentioned in item 5 of this list above can be used in a \texttt{CorrelatedGaussianObservables} set. Also, \texttt{BinnedObservables} or \texttt{FunctionObservables} can be used instead of and alongside \texttt{Observable}. If \texttt{noweight} is specified for any \texttt{Observable} then that particular \texttt{Observable} along with the corresponding row and column of the correlation matrix is excluded from the set of \texttt{CorrelatedGaussianObservables}.

Correlations between any set of observables can be computed using the construction:
\begin{lstlisting}[language=]
CorrelatedObservables <name> Nobs
Observable <name> <obs label> <histolabel> <min> <max> noMCMC noweight    <central value> <Gaussian error> <flat error>
Observable <name> <obs label> <histolabel> <min> <max> noMCMC noweight    <central value> <Gaussian error> <flat error>
...
...
<Total of Nobs lines of Observables>
<Nobs%*$\times$Nobs correlation matrix>
\end{lstlisting}
This prints the correlation matrix in the \texttt{Observables/Statistics.txt} file. All rules that apply to \texttt{CorrelatedGaussianObservables} also apply to \texttt{CorrelatedObservables}.

In addition, the inverse covariance matrix of a set of \texttt{Nobs Observables} can be specified with the following:
\begin{lstlisting}[language=]
ObservablesWithCovarianceInverse <name> Nobs
Observable <name> <obs label> <histolabel> <min> <max> MCMC weight    <central value> 0. 0.
Observable <name> <obs label> <histolabel> <min> <max> MCMC weight    <central value> 0. 0.
...
...
<Total of Nobs lines of Observables>
<Nobs%*$\times$Nobs inverse covariance matrix>
\end{lstlisting}

\item  {\bf A correlation between two observables} can be obtained with any of the four following specifications: 

\begin{lstlisting}[language=]
Observable2D <name> <obs1 label> <histolabel1> <min1> <max1> (no)MCMC (no)weight <obs2 label> <histolabel2> <min2> <max2>
#
Observable2D <name> <obs1 label> <histolabel1> <min1> <max1> MCMC file <filename> <histoname> <obs2 label> <histolabel2> <min2> <max2>
#
Observable2D <name> (no)MCMC (no)weight
(Binned)Observable <obs label 1> <histolabel 1> <min> <max> <central value> <Gaussian error> <flat error> (<bin_min> <bin_max>)
(Binned)Observable <obs label 2> <histolabel 2> <min> <max> <central value> <Gaussian error> <flat error> (<bin_min> <bin_max>)
#
Observable2D <name> MCMC file filename histoname
(Binned)Observable <obs label 1> <histolabel 1> <min> <max> (<bin_min> <bin_max>)
(Binned)Observable <obs label 2> <histolabel 2> <min> <max> (<bin_min> <bin_max>)
\end{lstlisting}

\item  {\bf Include configuration files} with the \texttt{IncludeFile} directive. This is useful if one
   wants to separate the input configurations for better organization and flexibility.

\end{enumerate}
{\bf \large Step 2: Monte Carlo configuration file:}\\

The parameters and options of the Monte Carlo run are specified in a
configuration file, separate from the one(s) for the model. Each line
in the file has a pair of a label and its value, separated by space(s)
or tab(s). The available
parameters and options are: \\\\
{\bf \texttt{NChains}}: The number of chains in the Monte Carlo run. A minimum of 5 is suggested (default). 
If the theory space is complicated and/or the number of parameters is
large then more chains
are necessary. The amount of statistics collected in the main run is proportional to the
number of chains.\\\\
{\bf \texttt{PrerunMaxIter}} : The maximum number of iterations that the pre-run will go through (Default: 1000000). 
The pre-run ends automatically when the chains converge (by default R$<$1.1, see below) and all efficiencies 
are adjusted. While it is not necessary for the pre-run to converge for a run to be 
completed, one should exercise caution if convergence is not attained.\\\\
{ \bf \texttt{NIterationsUpdateMax}}: The maximum number of iterations
after which the proposal functions are updated in the
pre-run and convergence is checked. (Default: 1000)\\\\ 
{\bf \texttt{Seed}}: The seed can be fixed for deterministic
runs. (Default: 0, corresponding to a random seed initialization)\\\\
{\bf \texttt{Iterations}}: The number of iterations in the main run. This run is for the purpose of
collecting statistics and is at the users discretion. (Default: 100000)\\\\
{\bf \texttt{MinimumEfficiency}}: This allows setting the minimum efficiency of 
all the parameters to be attained in the pre-run. (Default: 0.15)\\\\
{\bf \texttt{MaximumEfficiency}}: This allows setting the maximum efficiency of 
all the parameters to be attained in the pre-run. (Default: 0.35)\\\\
{\bf \texttt{RValueForConvergence}}: The $R$-value for which
convergence is considered to be attained in the pre-run can be set with this flag. (Default: 1.1)\\\\
{\bf \texttt{WriteParametersChains}}: The chains will be written in the \ROOT file MCout*.root. 
This can be used for analyzing the performance of the chains and/or to
use the sampled pdf for post-processing. (Default: false)\\\\
{\bf \texttt{FindModeWithMinuit}}: To find the global mode with \texttt{MINUIT} starting
from the best fit parameters in the MCMC run. (Default: false)\\\\
{\bf \texttt{RunMinuitOnly}}: To run a \texttt{MINUIT} minimization only, without running the MCMC. (Default: false)\\\\
{\bf \texttt{CalculateNormalization}}: Whether the normalization of
the posterior pdf will be calculated
at the end of the Monte Carlo run. This is useful for model comparison. (Default: false)\\\\
{\bf \texttt{NIterationNormalizationMC}}: The maximum number of iterations used to compute the normalization. (Default: 1000000)\\\\
{\bf \texttt{PrintAllMarginalized}}: Whether all marginalized distributions will be printed in a pdf file
(MonteCarlo\_plots\_*.pdf). (Default: true)\\\\
{\bf \texttt{PrintCorrelationMatrix}}: Whether the parametric correlation will be printed in ParamCorrelations*.pdf
and ParamCorrelations*.tex. (Default: false)\\\\
{\bf \texttt{PrintKnowledgeUpdatePlots}}: Whether comparison between prior and posterior knowledge will be 
printed in a plot stored in ParamUpdate*.pdf. (Default: false)\\\\
{\bf \texttt{PrintParameterPlot}}: Whether a summary of the parameters will be printed in ParamSummary*.pdf. (Default: false)\\\\
{\bf \texttt{PrintTrianglePlot}}: Whether a triangle plot of the parameters will be printed. (Default: false)\\\\
{\bf \texttt{WritePreRunData}}: Whether the pre-run data is written to
a file. Useful to exploit a successful pre-run for multiple runs. (Default: false) \\\\
{\bf \texttt{ReadPreRunData}}: Whether the pre-run data will be read from a previously stored prerun file. (Name of the file, default: empty)\\\\
{\bf \texttt{MultivariateProposal}}: Whether the proposal function
will be multivariate or uncorrelated. (Default: true)\\\\
{\bf \texttt{Histogram1DSmooth}}: Sets the number of iterative smoothing of 1D histograms. (Default: 0)\\\\
{\bf \texttt{Histogram2DType}}: Sets the type of 2D histograms: 1001
$\to$ Lego (default), 101 $\to$ Filled, 1 $\to$ Contour.\\\\
{\bf \texttt{MCMCInitialPosition}}: The initial distribution of chains over the parameter space. (Options: Center, RandomPrior, RandomUniform (default))\\\\
{\bf \texttt{PrintLogo}}: Toggle the printing of the \HEPfit logo on the histograms. (Default: true)\\\\
{\bf \texttt{NoHistogramLegend}}: Toggle the printing of the histogram legend. (Default: false)\\\\
{\bf \texttt{PrintLoglikelihoodPlots}}: Whether to print the 2D histograms for the parameters vs. loglikelihood. (Default: false)\\\\
{\bf \texttt{WriteLogLikelihoodChain}}: Whether to write the value of log
likelihood in a chain. (Default: false)\\\\
{\bf \texttt{Histogram2DAlpha}}: Control the transparency of the 2D histograms. This does not work with all 2D histogram types. (Default: 1)\\\\
{\bf \texttt{NBinsHistogram1D}}: The number of bins in the 1D histograms. (Default: 100, 0 sets default)\\\\
{\bf \texttt{NBinsHistogram2D}}: The number of bins in the 2D histograms. (Default: 100, 0 sets default)\\\\
{\bf \texttt{InitialPositionAttemptLimit}}: The maximum number of
attempts made at getting a valid logarithm of the likelihood for all chains before the pre-run starts. (Default: 10000, 0 sets default)\\\\
{\bf \texttt{SignificantDigits}} The number of significant digits appearing in the Statistics file. (Default: computed based on individual results, 0 sets default)\\\\
{\bf \texttt{HistogramBufferSize}}: The memory allocated to each
histogram. Also determines the number of events collected before
setting automatically the histogram range. (Default: 100000)\\

\noindent For example, a {\bf Monte Carlo configuration file} is written as: 

\begin{lstlisting}[language=]
NChains                    10
PrerunMaxIter              50000
Iterations                 10000
#Seed                       1
PrintAllMarginalized       true
PrintCorrelationMatrix     true
PrintKnowledgeUpdatePlots  false
PrintParameterPlot         false
MultivariateProposal       true
\end{lstlisting}

\noindent where a '\#' can be placed at the beginning of each line to comment it out.\\\\
{\bf \large Step 3: Run}\\\\
\noindent{\bf Library mode with MCMC:  }An example can be found in \texttt{examples/MonteCarloMode}

\begin{lstlisting}[numbers=none,language=]
  $ cd examples/MonteCarloMode
  $ make
\end{lstlisting}

After creating the configuration files, run with the command:
\begin{lstlisting}[numbers=none,language=]
  $ ./analysis <model conf> <Monte Carlo conf>
\end{lstlisting}
{\bf Alternative: Run with MPI}\\ \HEPfit allows for parallel processing of the MCMC run and the observable computations.
To allow for this \HEPfit, and \BAT have to be compiled with MPI support as explained in Section~\ref{sec:Installation}. The command

\begin{lstlisting}[numbers=none,language=]
  $ mpiexec -n N ./analysis <model conf> <Monte Carlo conf>
\end{lstlisting}
will launch analysis on \texttt{N} thread/cores/processors depending on the smallest
processing unit of the hardware used. Our MPI implementation allows for runs on multi-threaded single processors as
well as clusters with MPI support.
{\bf NOTE:} Our MPI implementation of \HEPfit cannot be used with
\BAT compiled with the \texttt{--enable-parallel} option. It is
mandatory to use the MPI patched version of \BAT as explained
in the \href{https://hepfit.roma1.infn.it/doc/latest-release/index.html}{online documentation}.\\

%
%%%%%%%%%%%%%%%%%%%%%%%%%%%%%%%%%%%%%
\subsection{Event generation mode}
\label{sec:EGM}
%%%%%%%%%%%%%%%%%%%%%%%%%%%%%%%%%%%%%
Using the model configuration file used in the Monte Carlo mode, one
can obtain predictions of observables. An example can be found in \texttt{examples/EventGeneration} folder:

\begin{lstlisting}[numbers=none,language=]
  $ cd examples/EventGeneration
  $ make
\end{lstlisting}

After making the configuration files, run with the command:
\begin{lstlisting}[numbers=none,language=]
  $ ./analysis <model conf> <number of iterations> [output folder]
\end{lstlisting}

The \texttt{<number of iterations>} defines the number of random points in the parameter space that will be evaluated. Setting this to 0 gives the value of the observables at the central value of all the parameters. If the \texttt{[output folder]} is not specified everything is printed on the screen and no data is saved. Alternately, one can specify the output folder and the run will be saved if \texttt{<number of iterations>} $>$ 0. The output folder can be found in \texttt{./GeneratedEvents}. The structure of the output folder is as follows.

\noindent{\bf Output folder structure:}
\begin{itemize}
\item \texttt{CGO}: Contains any correlated Gaussian observables that might have been listed in the model configuration files.
\item \texttt{Observables}: Contains any observables that might have been listed in the model configuration files.
\item \texttt{Parameters}: Contains all the parameters that were varied in the model configuration files.
\item \texttt{Summary.txt}: Contains a list of the model used, the parameters varied, the observables computed and the number of events generated. This can be used, for example, to access all the files from a third party program.
\end{itemize}
The parameters and the observables are stored in the respective directories in files that are named after the same. For example, the parameter \texttt{lambda} will be saved in the file \texttt{lambda.txt} in the \texttt{Parameters} folder.

%
%%%%%%%%%%%%%%%%%%%%%%%%%%%%%%%%%%%%%
\subsection{Library mode without MCMC}
\label{sec:LIBMC}
%%%%%%%%%%%%%%%%%%%%%%%%%%%%%%%%%%%%%
The library mode allows for access to all the observables implemented in \HEPfit
without a Monte Carlo run. The users can specify a \texttt{Model} and vary \texttt{ModelParameters}
according to their own algorithm and get the corresponding predictions for the observables. This is made possible through:
\begin{itemize}
\item a combined library: \texttt{libHEPfit.a} (installed in \texttt{HEPFIT\_INSTALL\_DIR/lib}).
\item a combined header file: \texttt{HEPfit.h} (installed in \texttt{HEPFIT\_INSTALL\_DIR/include/HEPfit}).
\end{itemize}
The \HEPfit library allows for two different implementations of the access algorithm.

\noindent{\bf Non-Minimal Mode:}

In the non-minimal mode the user can use the \texttt{Model conf} file to pass the default value of
the model parameters. The following elements must be present in the user code to define
the parameters and access the observable. (For details of model
parameters, observables, etc.~please look up the
\href{https://hepfit.roma1.infn.it/doc/latest-release/index.html}{online documentation}.)

\begin{lstlisting}[language=C++]
// Include the necessary header file.
#include <HEPfit.h>

// Define the model configuration file.
std::string ModelConf = "SomeModel.conf";

// Define a map for the observables.
std::map<std::string, double> DObs;

// Define a map for the parameters to be varied.
std::map<std::string, double> DPars;

// Initialize the observables to be returned.
DObs["Mw"] = 0;
DObs["GammaZ"] = 0.;
DObs["AFBbottom"] = 0.;

// Create and object of the class ComputeObservables.
ComputeObservables CO(ModelConf, DObs);

// Vary the parameters that need to be varied in the analysis.
DPars["Mz"] = 91.1875;
DPars["AlsMz"] = 0.1184;

// Get the map of observables with the parameter values defined above.
DObs = CO.compute(DPars);
\end{lstlisting}

\noindent{\bf Minimal Mode:}

In the minimal mode the user can use the default values in the \texttt{InputParameters} header file to define the
default values of the model parameters, therefore not requiring any additional input files to be
parsed. (For details of model name, flags, parameters, observables, etc.~please look up the
\href{https://hepfit.roma1.infn.it/doc/latest-release/index.html}{online documentation}.)

\begin{lstlisting}[language=C++]
// Include the necessary header file.
#include <HEPfit.h>

// Define a map for the observables.
std::map<std::string, double> DObs;

// Define a map for the mandatory model parameters used during initializing a model.
std::map<std::string, double> DPars_IN;

// Define a map for the parameters to be varied.
std::map<std::string, double> DPars;

// Define a map for the model flags.
std::map<std::string, std::string> DFlags;

// Define the name of the model to be used.
std::string ModelName = "NPZbbbar";

// Create and object of the class InputParameters.
InputParameters IP;

// Read a map for the mandatory model parameters. (Default values in InputParameters.h)
DPars_IN = IP.getInputParameters(ModelName);

// Change the default values of the mandatory model parameters if necessary.
// This can also be done with Dpars after creating an object of ComputeObservables
DPars_IN["mcharm"] = 1.3;
DPars_IN["mub"] = 4.2;

// Initialize the observables to be returned.
DObs["Mw"] = 0;
DObs["GammaZ"] = 0.;
DObs["AFBbottom"] = 0.;

// Initialize the model flags to be set.
DFlags["NPZbbbarLR"] = "TRUE";

// Create and object of the class ComputeObservables.
ComputeObservables CO(ModelName, DPars_IN, DObs);

// Set the flags for the model being used, if necessary.
CO.setFlags(DFlags);

// Vary the parameters that need to be varied in the analysis.
DPars["mtop"] = 170.0;
DPars["mHl"] = 126.0;

// Get the map of observables with the parameter values defined above.
DObs = CO.compute(DPars);
\end{lstlisting}
{\bf Use of hepfit-config:} A hepfit-config script can be found in the
\texttt{HEPFIT\_INSTALL\_DIR/bin} directory, which can be invoked with the 
following options:

\begin{lstlisting}[numbers=none,language=]
Library and Library Path: hepfit-config --libs

Include Path: hepfit-config --cflags
\end{lstlisting}

% The last command lists all the mandatory parameters in all the models sorted alphabetically and their
% default values as set in the class InputParameters.

%%%%%%%%%%%%%%%%%%%%%%%%%%%%%%%%%%%%%
\subsection{Custom models and observables}
\label{sec:custom}
%%%%%%%%%%%%%%%%%%%%%%%%%%%%%%%%%%%%%
A very useful feature of \HEPfit is that it allows the user to create
custom models and observables. We have already provided a template
that can be found in the \texttt{examples/myModel} directory which can
be used as a starting point. Below we describe how to implement both custom
models and custom observables.\\

\noindent {\bf Custom Models:}
The idea of a custom model is to define an additional set of
parameters over and above what is defined in any model in
\HEPfit. Typically the starting point is the \texttt{StandardModel}, as
in the template present in the \HEPfit package. Going by this template
in the \texttt{examples/myModel} directory, to create a model one has
to define the following:
\begin{itemize}
    \item In the \texttt{myModel.h} header file:
    \begin{enumerate}
        \item Define the number of additional parameters:
\begin{lstlisting}[numbers=none]
static const int NmyModelvars = 4;
\end{lstlisting}
        \item Define the variables corresponding to the parameters:
\begin{lstlisting}[numbers=none]
double c1, c2, c3, c4;
\end{lstlisting}
        \item Define getters for all the parameters:
\begin{lstlisting}[numbers=none]
double getc1() const { return c1; }
double getc2() const { return c2; }
double getc3() const { return c3; }
double getc4() const { return c4; }
\end{lstlisting}        
    \end{enumerate}
    \item In the \texttt{myModel.cpp} file:
    \begin{enumerate}
        \item Define the names of the parameters (they can be different from the variable names):
\begin{lstlisting}[numbers=none]
const std::string myModel::myModelvars[NmyModelvars] = {"c1", "c2", "c3", "c4"};
\end{lstlisting}
        \item Link the parameter name to the variable containing it for all the parameters:
\begin{lstlisting}[numbers=none]
ModelParamMap.insert(std::make_pair("c1", std::cref(c1)));
ModelParamMap.insert(std::make_pair("c2", std::cref(c2)));
ModelParamMap.insert(std::make_pair("c3", std::cref(c3)));
ModelParamMap.insert(std::make_pair("c4", std::cref(c4)));
\end{lstlisting}
        \item Link the names of the parameters to the corresponding variables in the \texttt{setParameter} method:
\begin{lstlisting}[numbers=none]
if(name.compare("c1") == 0)
        c1 = value;
    else if(name.compare("c2") == 0)
        c2 = value;
    else if(name.compare("c3") == 0)
        c3 = value;
    else if(name.compare("c4") == 0)
        c4 = value;
    else
        StandardModel::setParameter(name,value);
\end{lstlisting}        
    \end{enumerate}
\end{itemize}

This completes the definition of the model. One can also define flags that will control certain aspects of the model, but since this is an advanced and not so commonly used feature we will not describe it here. There is an implementation in the template for the user to follow should it be needed. Finally, the custom model needs to be added with a name to the \texttt{ModelFactory} in the main function as is done in \texttt{examples/myModel/myModel\_MCMC.cpp}.

\begin{lstlisting}[numbers=none]
ModelF.addModelToFactory("myModel", boost::factory<myModel*>() );
\end{lstlisting}

\bigskip
\noindent{\bf Custom Observables}

The definition of custom observables does not depend on having defined
a custom model or not. A custom observable can be any observable that
has not been defined in \HEPfit. It can be a function of parameters
already defined in a \HEPfit model or in a custom model or a
combination of the two. However, a custom observable has to be
explicitly added to the \texttt{ThObsFactory} in the main function as
is done in \texttt{examples/myModel/myModel\_MCMC.cpp}.
\begin{lstlisting}[numbers=none]
ThObsF.addObsToFactory("BIN1", boost::bind(boost::factory<yield*>(), _1, 1) );
ThObsF.addObsToFactory("BIN2", boost::bind(boost::factory<yield*>(), _1, 2) );
ThObsF.addObsToFactory("BIN3", boost::bind(boost::factory<yield*>(), _1, 3) );
ThObsF.addObsToFactory("BIN4", boost::bind(boost::factory<yield*>(), _1, 4) );
ThObsF.addObsToFactory("BIN5", boost::bind(boost::factory<yield*>(), _1, 5) );
ThObsF.addObsToFactory("BIN6", boost::bind(boost::factory<yield*>(), _1, 6) );
ThObsF.addObsToFactory("C_3", boost::factory<C_3*>() );
ThObsF.addObsToFactory("C_4", boost::factory<C_4*>() );
\end{lstlisting}
The first 6 observables require an argument and hence needed
\texttt{boost::bind}. The last two do not need an argument. The
implementation of these observables can be found in
\texttt{examples/myModel}
\texttt{/src/myObservables.cpp} and the corresponding
header file. In this template the \texttt{myObservables} class
inherits from the \texttt{THObservable} class and the observables
called \texttt{yield}, \texttt{C\_3} and \texttt{C\_4} inherit from
the former. Passing an object of the \texttt{StandardModel} class as a
reference is mandatory as is the overloading of the
\texttt{computeThValue} method by the custom observables, which is used to compute the value of the observable at each iteration.

%%%%%%%%%%%%%%%%%%%%%%%%%%%%%%%%%%%%%
\subsection{Example run in the Monte Carlo Mode}
\label{sec:example}
%%%%%%%%%%%%%%%%%%%%%%%%%%%%%%%%%%%%%

In this section we give an example of how \HEPfit can be used for a fit to data using the MCMC implementation in \BAT. Once you have installed \HEPfit following the instructions in Section~\ref{sec:Installation} move to the \texttt{MonteCarloMode} directory and compile the code with

\begin{lstlisting}[numbers=none,language=]
  $ cd examples/MonteCarloMode
  $ make
\end{lstlisting}
An example set of configuration files are packaged with the \HEPfit distribution. They can be found in the \texttt{examples/config} directory. For convenience we will copy this directory into the \texttt{MonteCarloMode} directory:

\begin{lstlisting}[numbers=none,language=]
  $ cp -r ../config .
\end{lstlisting}
The configuration files in that directory contain an example of a Unitarity triangle fit that can be done with experimental and lattice inputs. There are two files in the \texttt{config} directory:

\noindent\texttt{StandardModel.conf}: This file is the starting point of the model configurations for this example. It contains the definition of the model at the top. It then includes any other configuration files necessary for this example and a list of parameters that are mandatory for the SM implementation in \HEPfit. Note that all the parameters that are mandatory for \texttt{StandardModel} need not be present in this file but can also be present in any other configuration file that is included with the \texttt{IncludeFile} directive. The \texttt{StandardModel.conf} file looks like

\lstset{language=}
\begin{lstlisting}[numbers=none]
StandardModel
######################################################################
# Mandatory configuration files 
#---------------------------------------------------------------------
IncludeFile UTfit.conf
#
######################################################################
# Optional configuration files 
#---------------------------------------------------------------------
# IncludeFile Observables.conf
#
######################################################################
# Model Parameters 
#               name        ave         errg        errf
#---------------------------------------------------------------------
### Parameters in StandardModel
ModelParameter  GF        1.1663787e-5  0.          0.
# alpha=1/137.035999074
ModelParameter  ale   7.2973525698e-3   0.          0.
ModelParameter  AlsMz       0.118       0.0009      0.
ModelParameter  dAle5Mz     0.02750     0.00033     0.
ModelParameter  Mz          91.1875     0.0021      0.
ModelParameter  delMw       0.          0.          0.
ModelParameter  delSin2th_l 0.          0.          0.
\end{lstlisting}

\noindent\texttt{UTfit.conf}: This is the second file in the \texttt{config} directory and included from the \texttt{StandardModel.conf} file. This file contains the parameters that are relevant for a Unitarity Triangle fit and a list of \texttt{Observables} and \texttt{Observables2D} that are used in the fit. There are also some \texttt{ModelFlag} specifications in the file which determine the model specific run configurations. More details for these can be found in the \href{https://hepfit.roma1.infn.it/doc/latest-release/index.html}{online documentation}. The list of parameters looks similar to the one in \texttt{StandardModel.conf}

\begin{lstlisting}[numbers=none]
ModelFlag FlagCsi false
ModelFlag Wolfenstein false
######################################################################
# Model Parameters
#               name        ave         errg        errf
#---------------------------------------------------------------------
### Parameters for Flavour (Mandatory for all models)
#   scheme for bag parameters [NDR=0, HV=1, LRI=2]
# ModelParameter  lambda      0.2         0.          0.1
# ModelParameter  A           0.8         0.          0.3
# ModelParameter  rhob        0.0         0.          1.0
# ModelParameter  etab        0.0         0.          1.0
ModelParameter  V_us        0.22514     0.00055      0.
ModelParameter  V_cb        0.04045     0.00        0.01
ModelParameter  V_ub        0.00372     0.00023     0.
ModelParameter  gamma       1.22173     0.07        0.
\end{lstlisting}
while the list of observables looks like:
\begin{lstlisting}[numbers=none]
Observable  MtMSbar  MtMSbar  MtMSbar      1. -1.  noMCMC noweight
Observable  Dmd      DmBd     #Deltam_{d}  1. -1.  MCMC weight 0.5064    0.0019    0.
Observable  Dms      DmBs     #Deltam_{s}  1. -1.  MCMC weight 17.757    0.021     0.
Observable  EpsilonK EpsilonK #epsilon_{K} 1. -1.  MCMC weight 0.00228   0.00011   0.
#
### CKM Elements
Observable  Vud_in   Vud      V_{ud}       1. -1.  MCMC   weight   0.97420   0.00021   0.
#
Observable  alpha    alpha    #alpha       1. -1.  MCMC   weight   93.3	     5.6	   0.
#
### S coefficient of JPsiK time-dependent CPA
Observable  SJPsiK   SJPsiK   S_{J/#PsiK}  1. -1.  noMCMC noweight
Observable  C2beta   C2beta   Cos2#beta	   1. -1.  MCMC   weight    0.87  0.11  0.
\end{lstlisting}

\begin{figure}[t!]
    \centering
    \includegraphics[width = 0.32\textwidth]{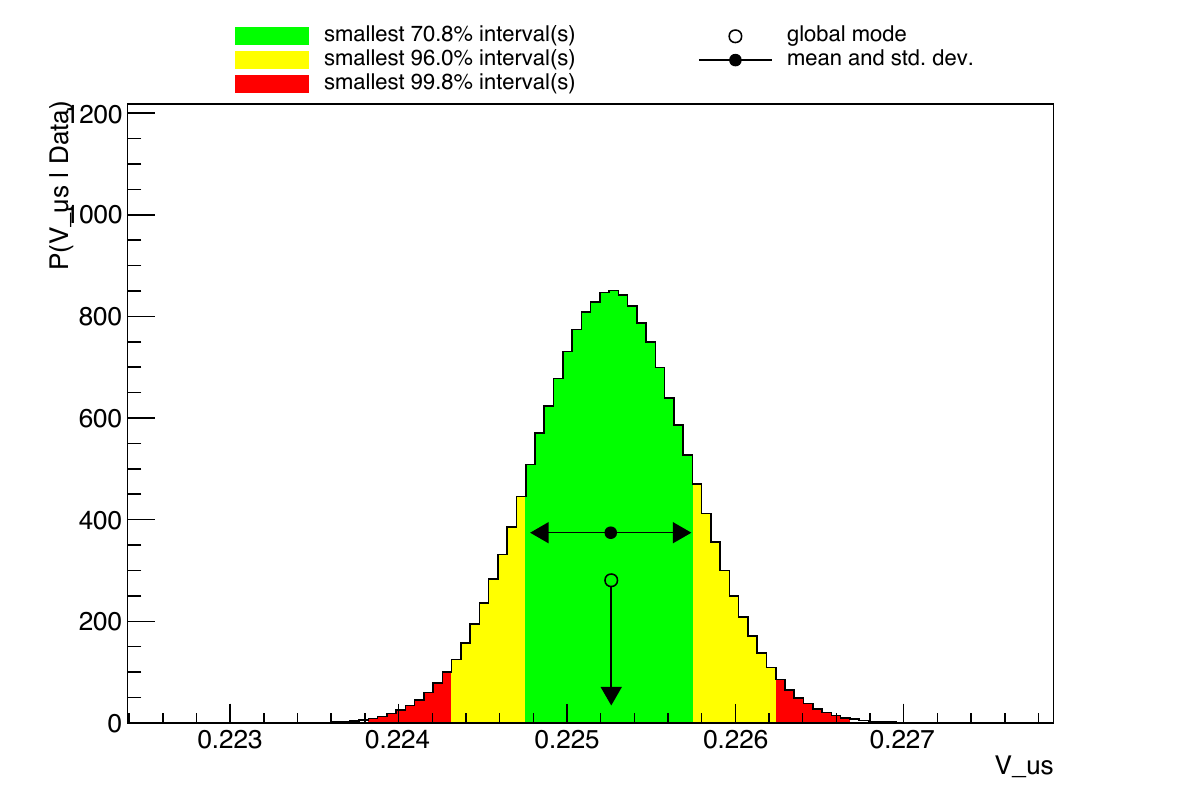}
    \includegraphics[width = 0.32\textwidth]{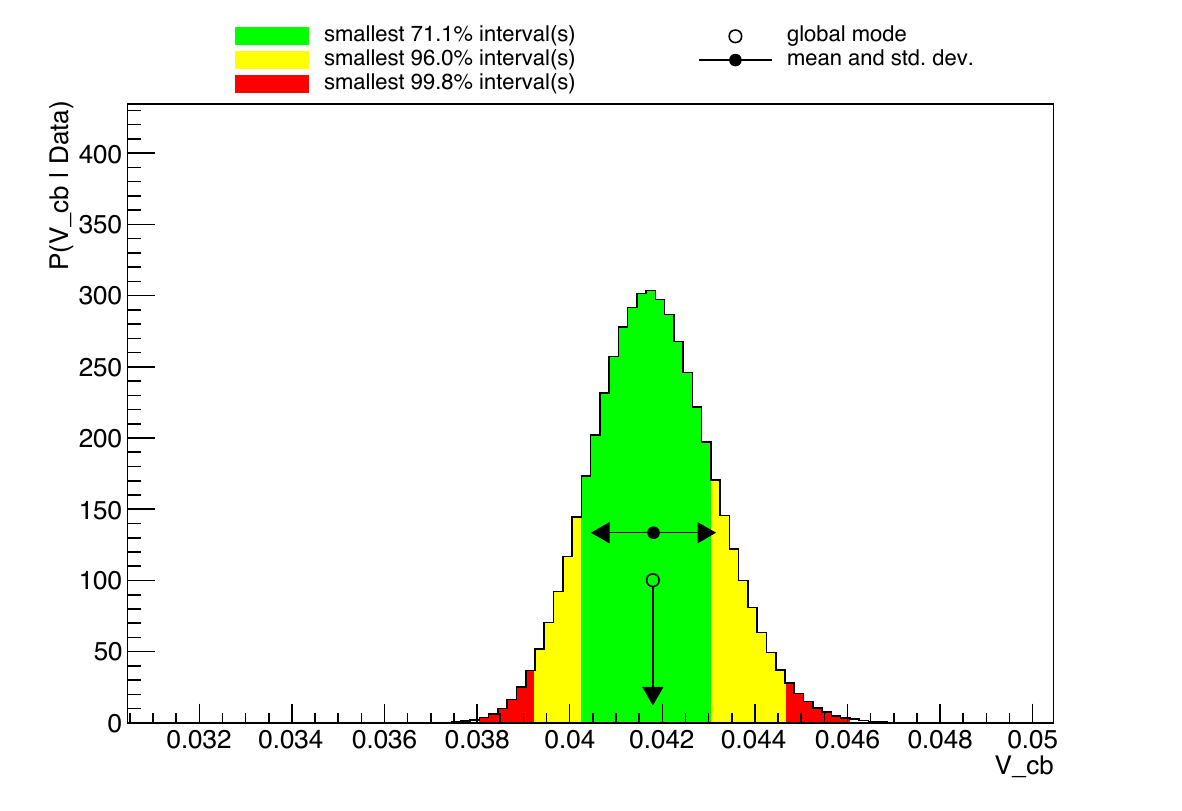}
    \includegraphics[width = 0.32\textwidth]{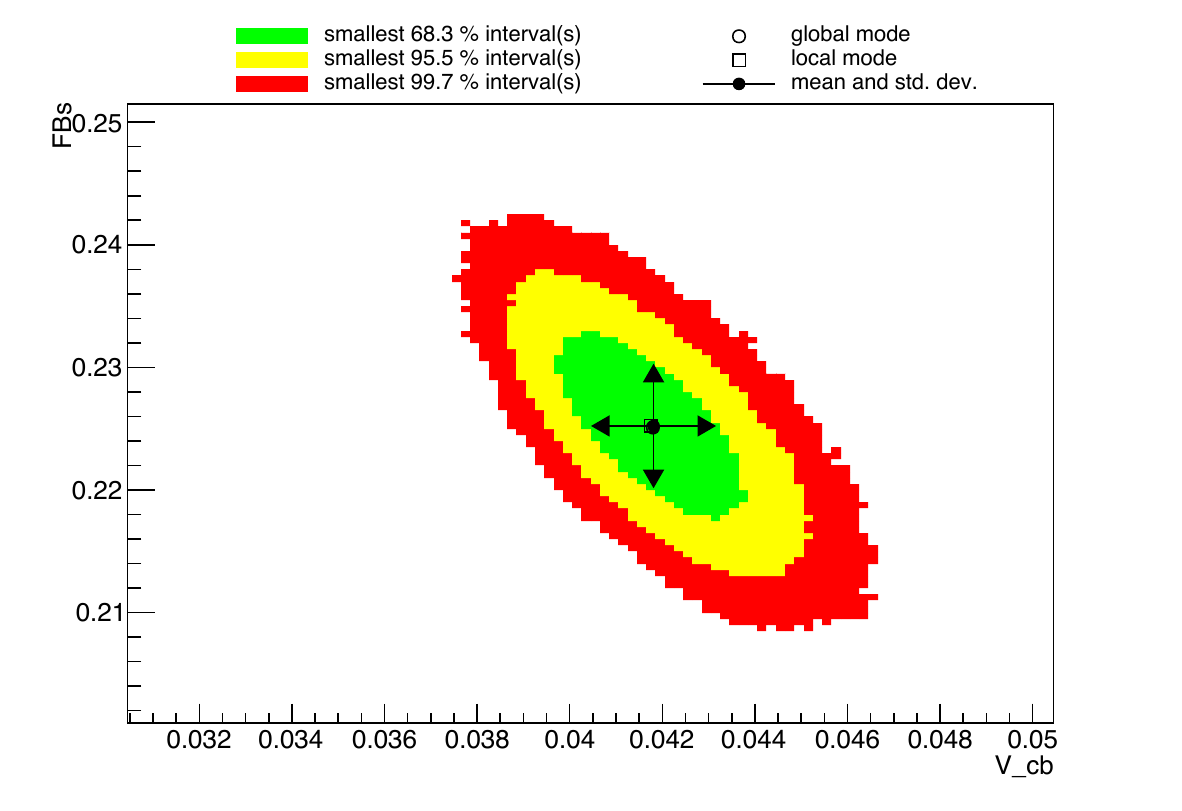}
    \caption{\it Example plots from a Unitarity Triangle fit that can be found in the \texttt{MonteCarlo\_plots.pdf} file.}
    \label{fig:MCP}
\end{figure}

\begin{figure}[t!]
    \centering
    \includegraphics[width = 0.32\textwidth]{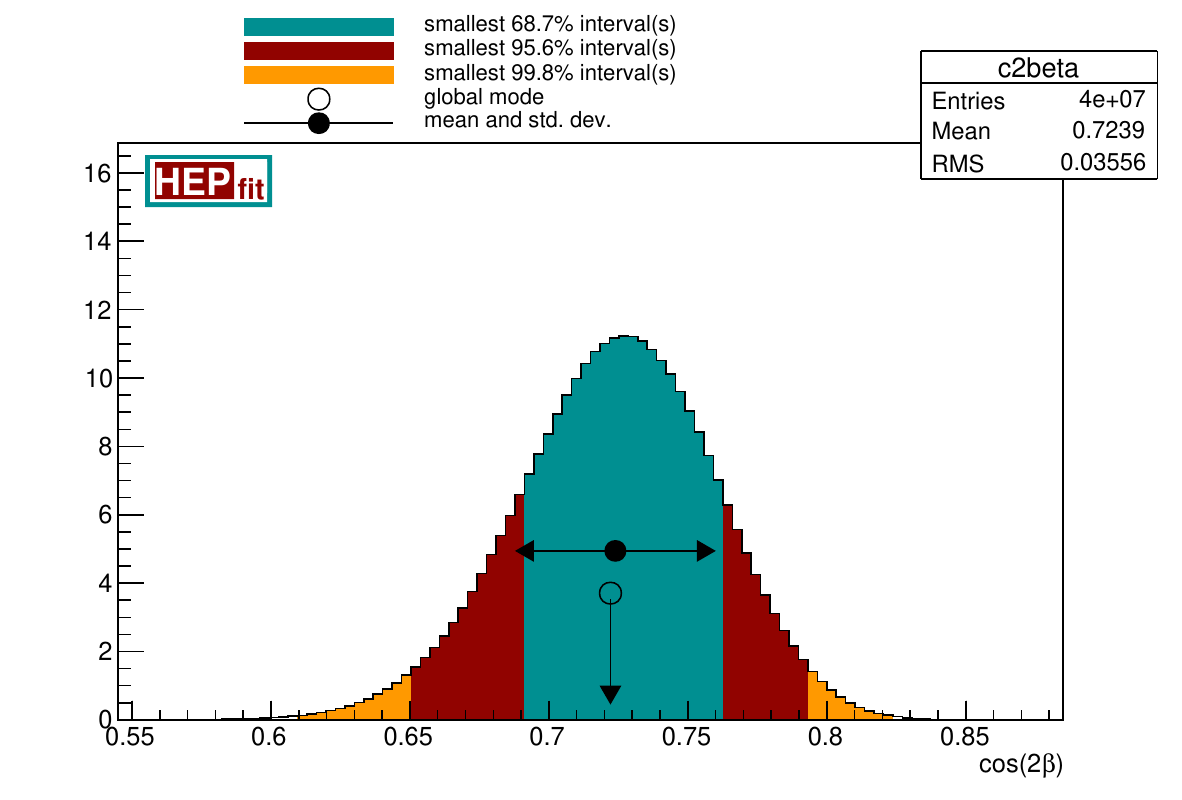}
    \includegraphics[width = 0.32\textwidth]{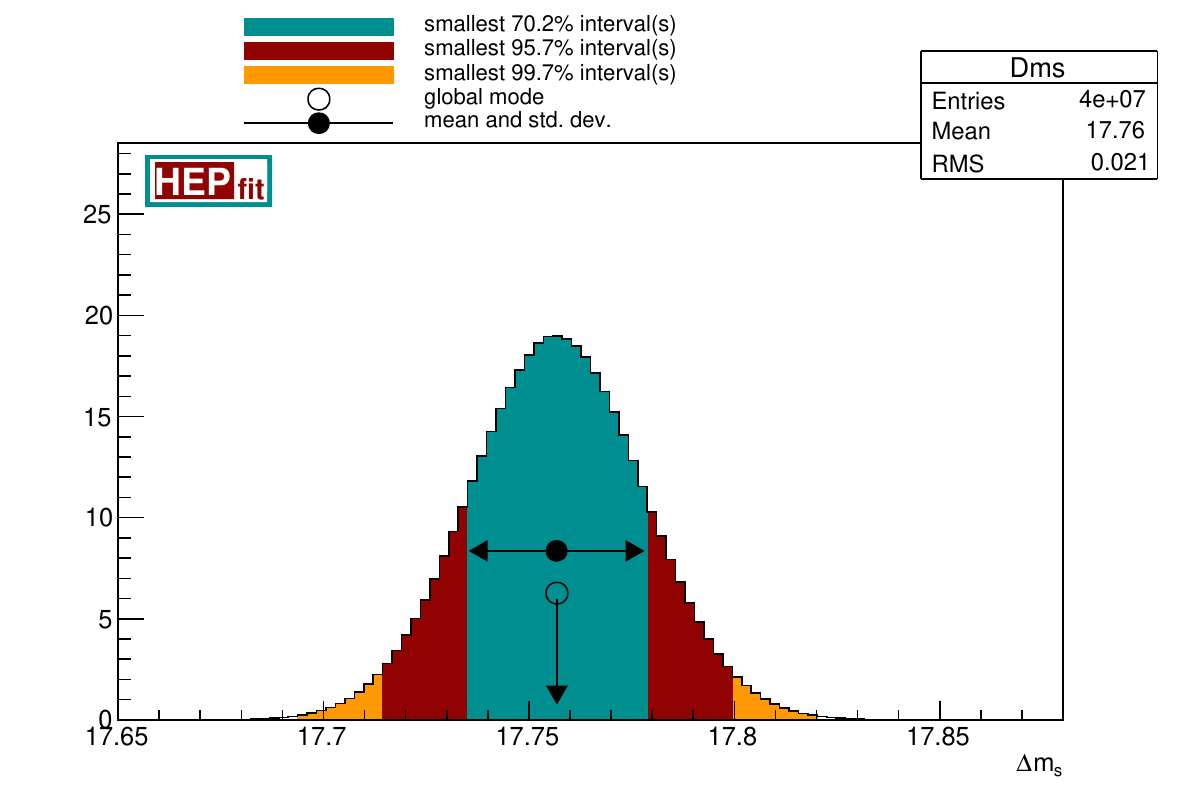}
    \includegraphics[width = 0.32\textwidth]{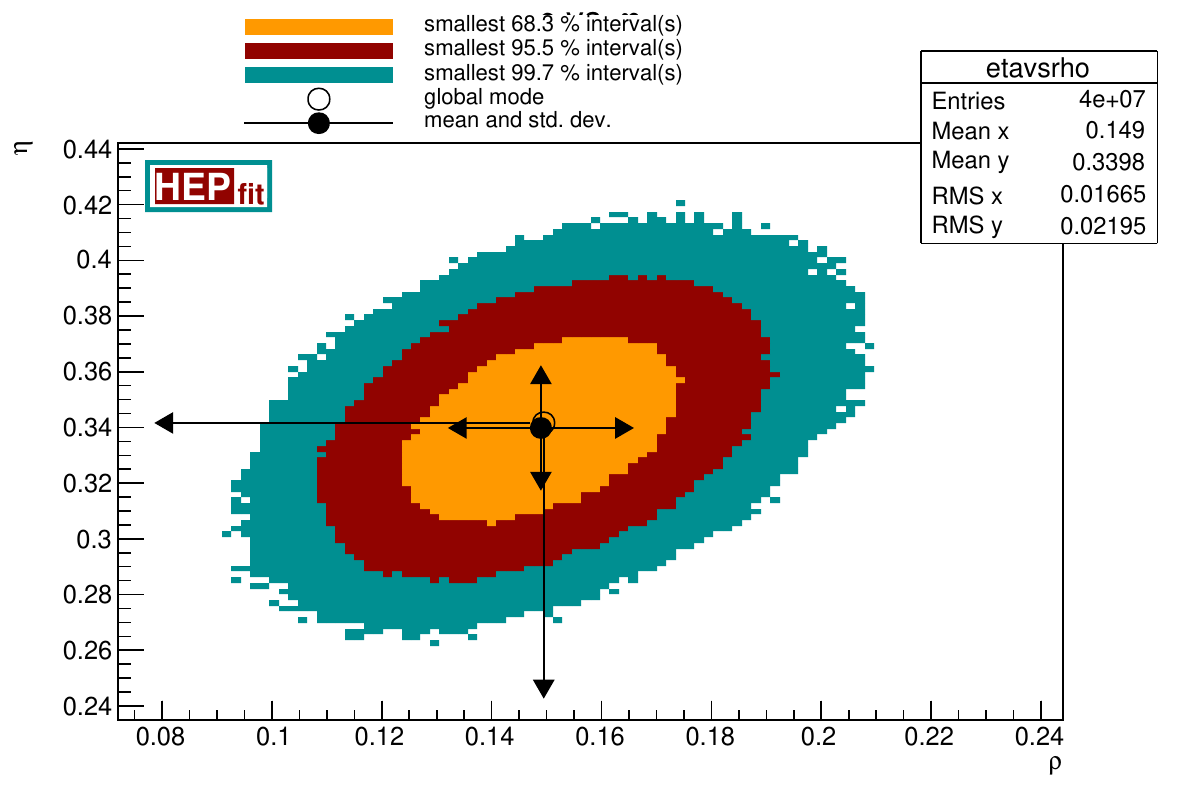}
    \caption{\it Example plots from a Unitarity Triangle fit that can be found in the \texttt{Observables} directory.}
    \label{fig:Observavbles}
\end{figure}

There is also a \texttt{MonteCarlo.conf} file in the \texttt{MonteCarloMode} directory. This file sets all the configurations of the MCMC run and can be used for any fit after any modifications that the user might choose to make. With these files a MCMC run can be started using the command:

\begin{lstlisting}[numbers=none]
  $ ./analysis config/StandardModel.conf MonteCarlo.conf
\end{lstlisting}
or
\begin{lstlisting}[numbers=none]
  $ mpiexec -N n ./analysis config/StandardModel.conf MonteCarlo.conf
\end{lstlisting}
where \texttt{n} is the number of CPU cores the user wants to use. We ran this fit with the following modifications to the \texttt{MonteCarlo.conf} file
\begin{lstlisting}[numbers=none]
## Number of chains
NChains                    40
## Max iterations in prerun
PrerunMaxIter              10000000
## Analysis iterations
Iterations                 1000000
\end{lstlisting}
Increasing the \texttt{PrerunMaxIter} allows for the convergence of the chains although, in this particular run convergence occurred at under 400,000 iterations. Increasing the \texttt{Iterations} allows for collection of moderate amount of statistics. In this configuration using 40 CPU cores, the fit took approximately 50 minutes to complete. The output generated by the code are:

\begin{itemize}
\item \texttt{log.txt}: The log file containing information on the MCMC run and is similar to the output at the terminal.
\item \texttt{MCout.root}: The \ROOT file containing all the information of the run and
  the histograms (and possibly the corresponding chains). This file can be read and all information processed using \ROOT.
\item \texttt{MonteCarlo\_results.txt}: A text file containing some
  information on the fitted parameters. Here one can find all the details of the distributions of the parameters like the mean, standard deviation, mode and percentiles. The snippet below shows an example:
  \begin{lstlisting}[numbers=none]
   Results of the marginalization
 ==============================
 List of variables and properties of the marginalized distributions:

  (0) Parameter "V_us" :
      Mean +- sqrt(Variance):         +0.22526 +- 0.0004694
      Median +- central 68% interval: +0.22526 + 0.000467903 - 0.000467598
      (Marginalized) mode:            +0.225278
       5% quantile:                   +0.224487
      10% quantile:                   +0.224658
      16% quantile:                   +0.224793
      84% quantile:                   +0.225728
      90% quantile:                   +0.225862
      95% quantile:                   +0.226034
      Smallest interval containing 68.0% and local mode:
      (0.22481, 0.225745) (local mode at 0.225278 with rel. height 1; rel. area 1)
  \end{lstlisting}
  
\item \texttt{MonteCarlo\_plots.pdf}: A file containing the 1D and 2D histograms for the parameters. Some example 1D and 2D histograms from the Unitarity Triangle fit can be found in Figure~\ref{fig:MCP}. The plots include approximate 68.3\%, 95.5\% and 99.7\% probability regions, computed from the mode of the 1D or 2D distributions, and some statistical information.

\item \texttt{Observables}: a directory containing the histograms for all the observables
  specified in the configuration file, as well as some text files. Some example plots from the Unitarity Triangle fit are shown in Figure~\ref{fig:Observavbles}.  

\item \texttt{Observables/HistoLog.txt}: A file containing the information on over-run and under-run
  during the filling of histograms.
\item \texttt{Observables/Statistics.txt}: A file containing a compilation of the statistics
  extracted from the histograms of the observables. For example:
\begin{lstlisting}[numbers=none]  
Observables:

  (4) Observable "EpsilonK":
      Mean +- sqrt(V):                0.0022802 +- 0.00010991
      (Marginalized) mode:            0.0022853
      Smallest interval(s) containing at least 69.2531% and local mode(s):
       (0.002168, 0.0023924) (local mode at 0.0022853 with rel. height 1; rel. area 1)

      Smallest interval(s) containing at least 95.8769% and local mode(s):
       (0.0020558, 0.0025046) (local mode at 0.0022853 with rel. height 1; rel. area 1)

      Smallest interval(s) containing at least 99.7487% and local mode(s):
       (0.0019436, 0.0026066) (local mode at 0.0022853 with rel. height 1; rel. area 1)
\end{lstlisting}       
It also contains some measures that can be used for judging the goodness of fit and models comparison. For details see~\cite{Ciuchini:2015qxb,Ciuchini:2017mik,doi:10.1080/01966324.2011.10737798}.
\begin{lstlisting}[numbers=none]
LogProbability at mode: 111.33
LogLikelihood at mode: -1.1484
LogAPrioriProbability at mode: 112.48


LogLikelihood mean value: -2.9074
LogLikelihood variance: 1.9054
IC value: 13.437
DIC value: 9.6258
\end{lstlisting}
\end{itemize}


Other files might be generated depending on the options specified in the Monte Carlo 
configuration file.

%%%%%%%%%%%%%%%%%%%%%%%%%%%%%%%%%%%%%%%
\section{Summary}
\label{sec:Sum}
%%%%%%%%%%%%%%%%%%%%%%%%%%%%%%%%%%%%%%%

\HEPfit is a multipurpose and flexible analysis framework that can be used for fitting models to experimental and theoretical constraints. It comes with the ability to use the Bayesian MCMC framework implemented in \BAT, which is highly efficient and allows for both factorized and non-factorized priors and is integrated with \ROOT. The key features of the \HEPfit framework are:

\begin{itemize}
    \item It allows for Bayesian analyses using an efficiently parallelized MCMC and for any other custom statistical analysis that the user might want to implement. This is made possible by allowing for the computation of the observables using the \HEPfit library.
    
    \item The Bayesian analysis framework in \HEPfit using \BAT is
      parallelized with MPI and can be run on a large number of
      processors without a substantial increase in the overhead. This
      makes the use of \HEPfit extremely scalable from desktop computers to large clusters. 
    
    \item Over and above the models and observables defined in
      \HEPfit, it also allows for users to define their own models and
      observables. This gives users the flexibility to use \HEPfit for
      any model and set of observables of their choice. User-defined models can add new parameters and the observables can be
      functions of these parameters and/or of the parameters already defined in \HEPfit. 
\end{itemize}

\HEPfit has been thoroughly tested over the years with several physics
results already published. This has allowed us not only to
gain confidence in the implementation of \HEPfit but also to gather
configuration files that are publicly available in the \HEPfit
repository and can be used by anyone wishing to start using \HEPfit
with minimal initial effort. In this article we give a summary of the structure of the code, the statistical framework used in \BAT, the parallelization of the code, a brief overview of the models and observables implemented in \HEPfit and an overview of the physics results that has been produced. 

Eventually, this article should be also retained as an optimal starting point for installing and running \HEPfit. For further technical details on the usage of \HEPfit and on the structure of the code, a comprehensive \href{https://hepfit.roma1.infn.it/doc/latest-release/index.html}{online documentation} is available on the \href{https://hepfit.roma1.infn.it/}{\HEPfit website}.

\acknowledgments

The \HEPfit developers would like to thank several physicists who
have contributed indirectly to its development with suggestions, bug
reporting and usage of the code. The developers would also like to
thank the authors of \texttt{flavio}~\cite{Straub:2018kue}, in particular David Straub, since
a large fraction of the flavour physics code in \HEPfit has been
tested against it. At various stages of its development, we have
compared \HEPfit with other publicly available codes, namely,
\texttt{FeynHiggs}~\cite{Heinemeyer:1998yj,Bahl:2018qog}, \texttt{SuperISO}~\cite{Mahmoudi:2008tp}, \texttt{SuSeFLAV}~\cite{Chowdhury:2011zr},
\texttt{Gfitter}~\cite{Flacher:2008zq}, \texttt{RunDec}~\cite{Herren:2017osy}, and \texttt{ZFITTER}~\cite{Akhundov:2013ons}. The
developers also acknowledge the use of \texttt{Madgraph}~\cite{Alwall:2014hca} and its
allied frameworks. L.R. has been supported by the U.S. Department of
Energy under grant DE-SC0010102. D.C. has been supported by the
(Indo-French) CEFIPRA/IFCPAR Project No.~5404-2.  Several of the
developers of \HEPfit were supported by the European Research Council
under the European Union’s Seventh Framework Programme
(FP/2007-2013)/ERC Grant Agreement n. 279972 ``NPFlavour''. This
project has received funding from the European Research Council (ERC)
under the European Union's Horizon 2020 research and innovation
program (grant agreement n$^\mathrm{o}$ 772369). M.C. is associated to the
Dipartimento di Matematica e Fisica, Universit{\`a} di Roma Tre, and
E.F. and L.S. are associated to the Dipartimento di Fisica,
Universit{\`a} di Roma ``La Sapienza''.  
V.M. and A.Pe. are supported by the Spanish Government and ERDF funds from the EU Commission [Grant FPA2017-84445-P], the Generalitat Valenciana [Grant Prometeo/2017/053], the Spanish Centro de Excelencia Severo Ochoa Programme [Grant SEV-2014-0398] and funded by Ministerio de Ciencia, Innovación y Universidades, Spain
[Grant FPU16/01911] and [Grant FPU15/05103]
respectively. M.F. acknowledge the financial support from MINECO
grant FPA2016- 76005-C2-1-P, Maria de Maetzu program grant
MDM-2014-0367 of ICCUB and 2017 SGR 929. G.G.d.C. is supported by the
National Science Centre, Poland, under research grant
no. 2017/26/D/ST2/00225. The work of S.M. has been supported by JSPS
KAKENHI Grant No.~17K05429. M.V. is supported by the NSF Grant
No.~PHY-1915005. N.Y. is supported by JSPS KAKENHI Grant Numbers JP15H05889, JP15K21733, and JP17H02875.
A.M.C. acknowledges support by the Swiss National Science Foundation (SNF) under contract 200021\_178967.
We would also like to thank the developers of \BAT, in particular F.~Beaujean, K.~Kr{\"o}ninger and D.~Greenwald for
support during various stages of integration of \BAT with \HEPfit.

\bibliographystyle{JHEP}
\bibliography{HEPfit}

\end{document}